\documentclass[journal]{IEEEtran}
\IEEEoverridecommandlockouts
\usepackage{cite}
\usepackage{amsmath,amssymb,amsfonts}
\usepackage{algorithm}
\usepackage{algpseudocode}
\usepackage{graphicx}
\usepackage{textcomp}
\usepackage{xcolor}
\usepackage{enumitem}
\usepackage{pifont}
\def\BibTeX{{\rm B\kern-.05em{\sc i\kern-.025em b}\kern-.08em
    T\kern-.1667em\lower.7ex\hbox{E}\kern-.125emX}}
\begin{document}

\title{Energy Management and Wake-up for IoT Networks Powered by Energy Harvesting}

\author{David E. Ru\'{i}z-Guirola, \IEEEmembership{Graduate Student Member, IEEE}, Samuel Montejo-S\'{a}nchez, \IEEEmembership{Senior Member, IEEE}, Israel Leyva-Mayorga, \IEEEmembership{Member, IEEE}, Zhu Han,  \IEEEmembership{Fellow, IEEE}, Petar Popovski, \IEEEmembership{Fellow, IEEE}, and Onel L. A. L\'{o}pez, \IEEEmembership{Senior Member, IEEE}
\vspace{-4mm}

\thanks{David E. Ru\'{i}z-Guirola and Onel L. A. L\'{o}pez are with the Centre for Wireless Communications, University of Oulu, Finland. \{David.RuizGuirola, Onel.AlcarazLopez\}@oulu.fi.  
Samuel Montejo-S\'{a}nchez is with the {Instituto Universitario de Investigaci\'{o}n y Desarrollo Tecnol\'{o}gico, Universidad Tecnol\'{o}gica Metropolitana}, Santiago, Chile. \{smontejo@utem.cl\}. Israel Leyva-Mayorga and Petar Popovski are with Department of Electronic Systems, {Aalborg University}, Aalborg, Denmark. \{ilm, petarp\}@es.aau.dk. Zhu Han is with the Department of Electrical {and Computer Engineering}, {University of Houston}, Houston TX, USA. \{zhan2@central.uh.edu\}.
}}

\maketitle

\begin{abstract}
The rapid growth of the Internet of Things (IoT) presents sustainability challenges such as increased maintenance requirements and overall higher energy consumption. This motivates self-sustainable IoT ecosystems based on Energy Harvesting (EH). This paper treats IoT deployments in which IoT devices (IoTDs) rely solely on EH to sense and transmit information about events/alarms to a base station (BS). The objective is to effectively manage the duty cycling of the IoTDs to prolong battery life and maximize the relevant data sent to the BS. Ths BS can also wake up specific IoTDs if extra information about an event is needed upon initial detection. We propose a K-nearest neighbors (KNN)-based duty cycling management to optimize energy efficiency and detection accuracy by considering spatial correlations among IoTDs’ activity and their EH process. We evaluate machine learning approaches, including reinforcement learning (RL) and decision transformers (DT), to maximize information captured from events while managing energy consumption. Significant improvements over the state-of-the-art approaches are obtained in terms of energy saving by all three proposals, KNN, RL, and DT. Moreover, the RL-based solution approaches the performance of a genie-aided benchmark as the number of IoTDs increases.
\end{abstract}


\section{Introduction}\label{intro}
 
Achieving sustainable development in today's society requires technological solutions that effectively balance economic growth, social equity, and environmental integrity~\cite{lopez2023energy}. 
These solutions must optimize resource management, improve access to essential services like healthcare and education, and minimize environmental impacts through energy-efficient practices and pollution monitoring. 
The Internet of Things (IoT) plays a pivotal role in this context by enabling connectivity across a wide range of applications, including mobility, safety, environmental monitoring, and resource management, to name a few~\cite{belli2020iot}. 
Nevertheless, ensuring sustainability and autonomy in IoT systems demands a reduction in dependency on external power sources. Energy harvesting (EH) emerges as a solution, targeting self-sustainability, in which the average energy consumed over a given period is offset by the battery level and the energy harvested during the same period~\cite{lopez2024zero}. Despite fluctuations in consumption and EH rates, a well-designed system can achieve long-term self-sustainability, eliminating the need for recharging or manual battery replacement and supporting sustained energy autonomy~\cite{lopez2024zero}. 

Emerging network architectures increasingly emphasize sustainability, particularly given the expected ultra-dense deployment of IoT devices (IoTDs). This vision necessitates both reducing energy consumption and efficient resource allocation to maintain scalability and reliability~\cite{lopez2023energy}. 
Techniques such as wake-up receiver (WuR) and discontinuous reception (DRX) can help in this regard. 
WuR, for example, can potentially save up to 1,000 times the energy consumed by the main radio frequency interface (RFI), allowing IoTDs to transition to a ``sleep'' state while the WuR remains active~\cite{ruiz2022energy}. 
This allows for on-demand operation, where the main RFI consumes energy only during communication~\cite{rostami2020novel}. 
However, configuring DRX and WuRs remains challenging due to the irregular traffic patterns in IoT networks~\cite{ruiz2022energy}. Additionally, in self-sustainable scenarios where IoT devices rely solely on ambient EH, fluctuating energy availability poses a risk to their ability to recharge and maintain continuous operation~\cite{lopez2024zero}. These uncertainties complicate the design of effective energy management and duty-cycling strategies, both essential for ensuring long-term network scalability and functionality. Thus, it becomes vital to develop energy-aware mechanisms that dynamically adapt IoTDs' operation modes (\textit{e.g.}, ``sleep'', ``idle'', and ``sensing''), thereby ensuring sustainable operation throughout the network’s lifetime~\cite{lopez2023energy}.

\vspace{-5ex}
{\begin{center}
    \begin{table*}[t]
    \caption{Applications/Examples of low-power IoT event-triggered traffic Scenarios}
    \label{IoT_cases}
    \centering
    \begin{tabular}{p{1.7cm}p{1.8cm}p{4.7cm}p{2.2cm}p{2cm}p{1.55cm}p{0.5cm}}
    \hline
    \textbf{Application} &   \textbf{Example} &   \textbf{Distinctive features} &   \textbf{Typical sensors} 
    &   \textbf{Commun Range}
    &   \textbf{Density} 
    &   \textbf{Ref.} \\  
    \hline
    Motion detection& Smart home security& Binary (motion/no motion) real-time detection, limited range and angle detection & Passive infrared  sensors& Short/Medium& Low/Medium &\cite{juan2016development}\\
    Fire alarm system& Building safety& High sensitivity, battery backup, placed on enclosed areas& Smoke detectors, heat sensors& Short& Medium/High &\cite{li2012effect}\\
    Environmental monitoring& Air quality& High durability and scalability & CO$_2$, NO$_2$, air quality sensors& Medium/Long& Low/Medium &\cite{realdata}\\
    Agricultural monitoring& Smart irrigation systems& Wireless control, weather resistance, automated systems& Soil moisture, temperature sensors& Medium/Long& Low &\cite{realdata}\\
    Health monitoring& Remote patient monitoring& Wearable sensors, secure data, real-time analytics& Heart rate, glucose sensors& Short& Medium/High &\cite{wearable}\\
    \hline
\end{tabular}
\vspace{-1ex}
\label{IoT_cases}
\end{table*}
\end{center}

In several IoT scenarios, a base station (BS) acts as a gateway for a group of IoTDs aiming to sense/detect events or alarms. These IoTDs are deployed to identify event triggers, playing a crucial role in applications such as motion/fire detection, environmental/agricultural monitoring and remote healthcare. These applications require real-time monitoring and immediate responses to changes in the environment or conditions, as outlined in Table~\ref{IoT_cases}. 
The IoTDs must meet the specific requirements of their target applications, ensuring reliable performance and long-term sustainability. Their deployment and management are thus essential for enhancing energy efficiency, safety, and sustainability across various sectors. However, as the number of deployed IoTDs increases, so do the challenges related to coordination, scalability, and energy efficiency, highlighting the need for intelligent coordination strategies and adaptive, energy-aware mechanisms~\cite{belmega2022online}. 

This motivates the development of novel frameworks addressing energy-efficient and sustainable operation in low-power IoT networks, particularly under event-driven (ED) traffic and energy constraints. Next, we review the state-of-the-art approaches and introduce our solution to these challenges.

\vspace{-1ex}
\subsection{State-of-the-Art}

Notably, the strategic deployment of IoTDs based on spatial correlation information can improve coverage, minimize blind spots, and optimize data transmission, increasing energy efficiency~\cite{TCOM_David}. 
Tools like Voronoi diagrams~\cite{Voronoi_ma2021escvad} and  clustering~\cite{ul2022learning} have been helpful for this. 
For instance, the authors in~\cite{raghuwanshi2023channel} propose a novel channel scheduling method by leveraging clustering-based techniques, and the spatial correlation between the device's activation. 
However, these tools alone cannot capture the sporadic traffic patterns of IoTDs~\cite{benkhelifa2020recycling}. 
To address this, recent studies have combined them with machine learning (ML) to characterize network performance, considering the spatial deployment correlation between the IoTDs generating the traffic~\cite{ul2022learning}. 
Additionally, recent advances in transformers present a promising advancement for IoT, offering superior sequential modeling capabilities and scalability~\cite{li2023survey}. Their integration with reinforcement learning (RL) enhances decision-making in resource-constrained environments like edge computing~\cite{hu2024transforming}, improving energy efficiency in IoT networks. 


ML techniques can facilitate the identification of IoT behavior patterns, enabling the implementation of energy-saving strategies such as sleep modes. 
Long-short term memory (LSTM)-based algorithms for learning traffic profiles, energy availability, and power consumption dynamics are presented in~\cite{ruiz2022energy}, lightening the computation/prediction-related tasks at low-power IoTDs. 
In particular, novel deep RL schemes are proposed in~\cite{salem2019traffic} to manage advanced sleep modes, 
where the BS sequentially determines the duration of the “sleep” state. 
Furthermore, in~\cite{managre2022review},  support vector machines (SVM), linear regression (LR), and neural networks (NN) are used to predict power consumption in smart grids, achieving up to 84\% accuracy. 
Similarly, \cite{ML_K3} employs a combination of decentralized and centralized RL to allocate spreading factor and transmission power, achieving significant improvements in network-level throughput and energy consumption, particularly in large and congested networks. 
Meanwhile, in~\cite{shabana2019energy}, almost an 80\% improvement in the performance of EH solutions for sustainable IoT systems is achieved by using RL, significantly reducing the energy consumption of IoTDs. 
Furthermore, EH efficiency is optimized in~\cite{nadali2023machine} by estimating the optimal location of non-static IoTDs using techniques such as K-nearest neighbors (KNN) and random forest. 

{Despite previous IoT-related research projects, challenges like scalability, resource efficiency, and energy efficiency persist in effectively implementing correlation-based strategies~\cite{belmega2022online}. 
Sustainable IoT development requires a holistic approach integrated with energy-efficient technologies, optimized resource allocation strategies, and ML-based solutions. 
Notably, research is missing on ML-based methods for adjusting WuR and duty cycling parameters in self-sustaining IoT networks, constituting an important research gap for further increasing energy efficiency that we have been exploring recently. 
} 
{Our preliminary work~\cite{ruiz2024intelligent} focuses on managing IoTD energy consumption and EH capabilities, extending the battery life of IoTDs, and addressing low-energy availability. 
Therein, we considered the scenario where the IoTDs gather ambient energy and monitor events and alarms while in the sensing state, and proposed a KNN-based duty cycling configuration. 
Additionally, a wake-up method was used to minimize the mean energy consumption of the IoTDs while considering the spatial correlations of their activities and a minimum information threshold required by the BS from events/alarms. The KNN-based configuration proposed in~\cite{ruiz2024intelligent} showed improvement in energy saving and performance with up to 11 times less misdetection probability and a 50\% decrease in energy consumption.  
However, while a fixed, predetermined information threshold may be practical for a specific setup, it is often impractical in broader contexts, as different IoT deployments face unique coverage and optimization challenges that require tailored solutions. 
Furthermore, it does not account for the inherent fluctuations and variability in the EH process, which could impact its effectiveness in real-world scenarios.}

\begin{table}[t]
    \caption{Brief Summary: State-of-the-Art vs. Our Proposal}
    \label{SoA_comparison}
    \centering
    \begin{tabular}{p{0.55cm}p{0.3cm}p{0.2cm}p{2.35cm}p{1.85cm}p{1.15cm}}
    \hline
    \textbf{Ref.} 
    & \textbf{WuR} 
    & \textbf{EH} 
    & \textbf{Goal} 
    & \textbf{Focus}  
    & \textbf{Tools} \\
    \hline
    \cite{ruiz2022energy} 
    & \centering{\checkmark} 
    & \centering{\ding{55}}
    & Minimize idle listening 
    & Traffic-based WuR control 
    & LSTM\\
    \cite{raghuwanshi2023channel} 
    & \centering{\ding{55}}  
    & \centering{\ding{55}} 
    & Reduce redundant channel access 
    & Channel scheduling   
    & Clustering\\
    \cite{salem2019traffic} 
    & \centering{\ding{55}}  
    & \centering{\ding{55}}
    & Optimize sleep mode to save energy 
    & Dynamic duty cycling   
    & Deep RL\\
    \cite{managre2022review} 
    & \centering{\ding{55}}  
    & \centering{\ding{55}} 
    & Optimize resource allocation 
    & Power prediction
    & NN, LR, SVM\\
    \cite{ML_K3} 
    & \centering{\ding{55}}  
    & \centering{\ding{55}}
    & Optimize adaptive resource control 
    & Spreading factor and transmission power
    & RL\\
    \cite{shabana2019energy} 
    & \centering{\ding{55}}  
    & \centering{\checkmark}
    & Efficient EH   
    & EH strategy 
    & RL\\
    \cite{nadali2023machine} 
    & \centering{\ding{55}}  
    & \centering{\checkmark}
    & Maximize EH collection 
    & Node placement for EH  
    & KNN, clustering \\
    \cite{ruiz2024intelligent} 
    & \centering{\checkmark}
    & \centering{\checkmark}
    & Optimize duty cycling 
    & Spatially aware duty cycling 
    & KNN \\
    {Our work} 
    & \centering{\checkmark}
    & \centering{\checkmark}  
    & Achieve self-sustainability and info gain 
    & Optimize jointly duty cycling, WuR, and ED
    & RL + KNN + clustering\\
    \hline
\end{tabular}
\vspace{-2ex}
\label{SoA_comparison}
\end{table}

\subsection{Contributions \& Organization of this work}

{In this paper, we build upon~\cite{ruiz2024intelligent}, using the framework and wake-up method proposed therein to maximize the amount of information the BS regarding events/alarms, rather than focusing solely on minimizing energy consumption for a predetermined minimum amount of information.  
By doing this, this work establishes a comprehensive framework balancing the information collected and the energy consumption required to achieve it, thus enhancing the overall energy efficiency and reliability of IoT networks in diverse applications. 
Table~\ref{SoA_comparison} summarizes the main features of the proposed solution and the state-of-the-art proposals.

The contributions of this paper are summarized as follows:}  
\begin{itemize}
    \item We introduce a BS‐controlled duty‐cycling scheme that combines KNN with Voronoi‐diagram analysis to dynamically determine, for each event, the minimum data each IoTD must report to maintain efficient operation, in contrast to the static configurations used in~\cite{ruiz2024intelligent}. This approach considers energy efficiency while learning the spatial correlations among IoTDs, enabling more intelligent scheduling of sensing and transmission activities. 
    \item We propose advanced RL-based approaches, including $Q$-learning and decision transformer (DT), to optimize duty cycling and wake-up methods. 
    These methods enable dynamic adaptation to varying network conditions, traffic patterns, and energy availability, making them well-suited for self-sustaining IoT environments. 
    \item We show that the proposed KNN and RL-based configurations achieve significant performance improvements, \textit{e.g.}, up to 4.5 times more information received per event and a 23\% decrease in energy consumption compared to a random duty cycling baseline. 
\end{itemize} 

\textcolor{black}{The rest of the paper is organized as follows. Section~\ref{system} presents the system model, including the models of EH, power consumption, and the influence of events on the IoTDs. 
Section~\ref{battery} describes the battery state evolution, while  Section~\ref{analisis} shows the wake-up method and the IoTD transition probabilities. Section~\ref{framework} outlines the optimization problem for the duty cycling and wake-up setup, and presents solutions based on heuristics, Voronoi diagrams, and ML. The effectiveness of the proposed methods is evaluated through numerical simulations in Section~\ref{result}. Finally, we conclude the paper in Section~\ref{conclusion}.}

\textit{Notation:} $(\cdot)^\mathrm{T}$ represents the transpose of a vector/matrix. Let $\boldsymbol{\mathbf{1}}$ denote a row vector of ones, $\mathbf{1}_0(\cdot)$ represent the indicator function, while $\lceil \cdot \rceil$ the ceiling function and $\beta_x(\cdot,\cdot)$ the regularized incomplete beta function. Moreover, we use \textbf{bold} capital/lowercase formatting to refer to matrices/vectors. 

\section{System Model}\label{system}

\begin{figure}[t!]
\centerline{\includegraphics[width=\columnwidth]{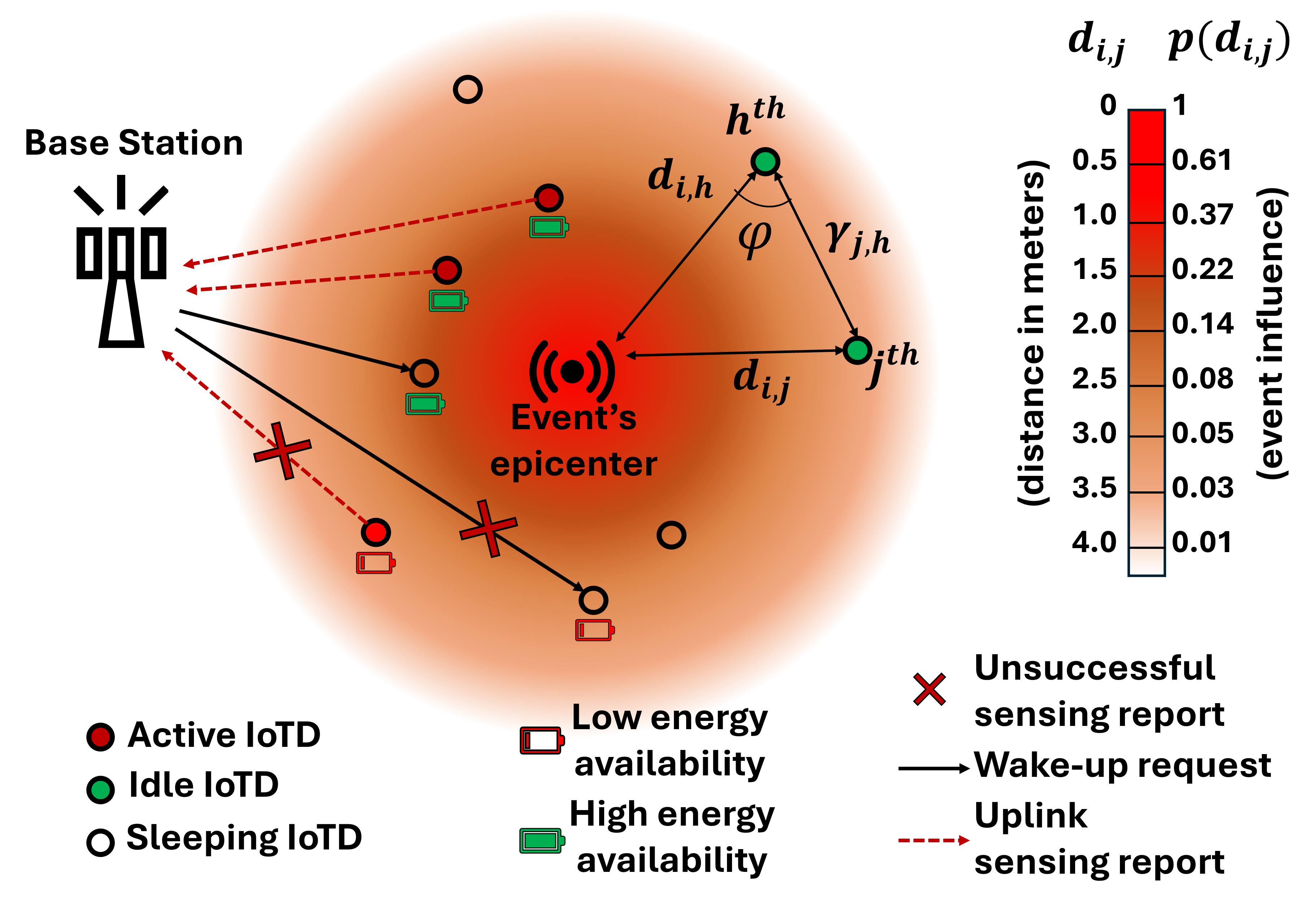}}
\vspace{-3mm}
\caption{Illustration of an IoT network where the BS collects information from various IoTDs that rely on EH to remain operational. The impact of an event on the surrounding IoTDs is modeled through a probability activation function that decays with the distance from the event epicenter to the IoTDs. The BS may wake-up certain IoTDs with the hope they can provide more information about a sensed event.}
\vspace{-2ex}
\label{fig1}
\end{figure}
We consider the coverage area of a BS that serves as the gateway for a set ${\mathcal{N}}$ of $N$ IoTDs with EH capabilities as depicted in Fig.~\ref{fig1}. Moreover, the time is slotted into transmission time intervals (TTIs) with duration $\tau$. 
We assume that the coordinator knows the position of each IoTD and that the epicenter of each event follows a two-dimensional uniform random distribution. Events occur independently in each TTI with probability~$\alpha$~\cite{ruiz2022energy}.

To describe the operation of an IoTD, let us consider four main modules, as depicted in~\figurename~\ref{IoTDModules}. The main RFI handles conventional data exchange with other IoTDs or the BS; the sensor detects event occurrences; the EH module harvests ambient energy to sustain operation; and the WuR module includes a low-power radio interface to detect wake-up signals and activate the main radio when needed.
An IoTD $j$ can operate in four different states ${{S^{(j)}\in {\mathcal{S}}}=\{S_1, S_2, S_3, S_4\}}$:
\begin{enumerate}
    \item {\em``idle'' state}, $S_1
    $, wherein it is waiting to be triggered by an event; 
    \item {\em``active'' state}, $S_2
    $, wherein the IoTD is triggered by an event and thus aims to send the information to the BS; 
    \item {\em``transmission'' state}, $S_3
    $, wherein it is sending the information to the BS through the main RFI; and
    \item {\em``sleep'' state}, $S_4
    $, wherein it is staying at a low-power consumption state with only the WuR operational, while the main RFI and sensor modules are turned off.  
\end{enumerate}
The WuR allows the BS to trigger the IoTD when additional information is needed, ensuring availability even during state $S_4$. 
IoTDs can harvest energy whenever it is available, regardless of their operational state.  
We denote $P_{m,n}^{(j)}$ as the transition probability for IoTD $j$ from a state $S_m$ into a state $S_n$.  
The states and their transition probabilities are described through a discrete-time Markov process, as shown in Fig.~\ref{fig2}.
\begin{figure}[t!]
\centerline{\includegraphics[width=\columnwidth]{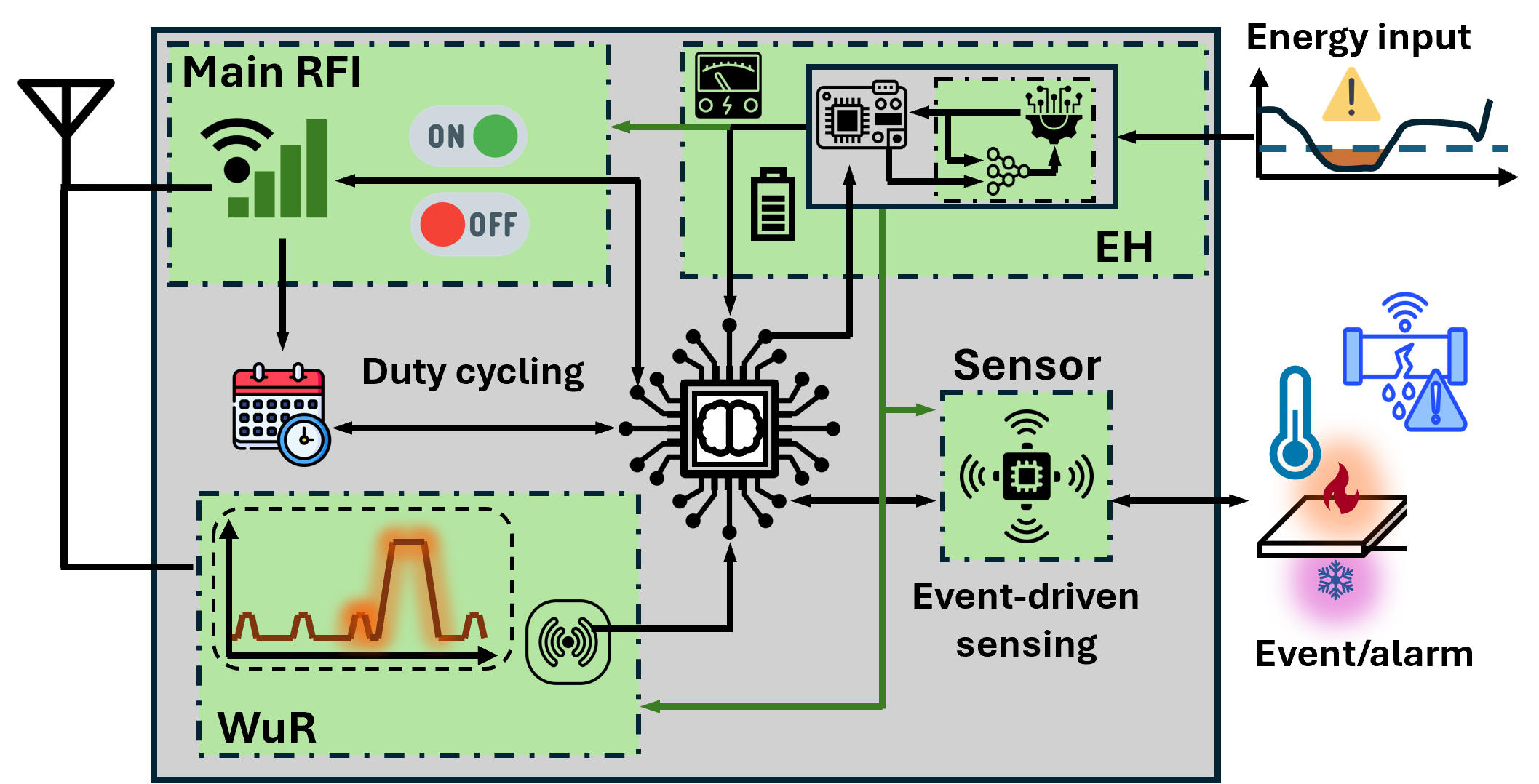}}
\vspace{-2mm}
\caption{Illustration of an IoTD architecture equipped with WuR and EH capabilities, showing its four main modules: main RFI, WuR, sensor, and EH unit. The EH circuit powers the other modules.}
\vspace{-3mm}
\label{IoTDModules}
\end{figure}

\subsection{Event Sensing and Information Level}

We define $p(d_{i,j})$ as the sensing function for an IoTD $j\in \mathcal{N}$ in the sensing state $S_1$. This function represents the impact of the $i^\text{th}$ event with epicenter at $(x_i, y_i)$ on the $j^{th}$ IoTD, within the network area $\xi \subset \Re^{2}$. 
Here, $d_{i,j}$ indicates the distance between the epicenter of the $i^{th}$ event and the $j^{th}$ IoTD. Note that, ${p(d_{i,j})
\rightarrow (0,1]}$ is non-increasing, modeling a decaying influence of events as the distance $d_{i,j}$ increases. Fig.~\ref{fig1} illustrates an example of the influence of an event epicenter on the surrounding IoTDs and note that the spatial activity correlation depends on $p(d_{i,j})$.  
Active devices transmit packets to the BS, managing all information exchanges within its cell~\cite{ruiz2022energy}, all with the same transmission power. 
In 
state $S_2$ (``active''), devices generate traffic, while no traffic is generated in the other states.

Active IoTDs send information about the detected event. 
In this study, we model the amount of information sent by each IoTD in state $S_2$ as 
\vspace{-1mm}
\begin{equation}
    I_{i,j} = \Psi p(d_{i,j}),
    \vspace{-1mm}
    \label{entropia}
\end{equation}
where $\Psi > 0$ constitutes the maximum amount of information that an IoTD can capture from an event. 
The BS gathers the information reported by the IoTDs. Specifically the information about the $i^\text{th}$ event at the BS is given by 
\begin{equation}
    I_{i} = 
    g(\{I_{i,j}\}_{\forall j: S^{(j)}=S_2}).
    \label{totalentropia}
\end{equation}
Here, $I_{i,j}=0$ for those IoTDs with insufficient energy to transmit information or for those where the event is outside the sensing range. Therefore, if no IoTD detects the event, then $I_{i} = 0$. 
Meanwhile, the function $g$ determines the amount of information that the BS can collect from the IoTDs triggered by events, and there may be several formulations for it. This function accounts for the correlation between the information collected by different devices from the same event. 

One potential function 
$g$ is $\max_{j}(I_{i,j})$, which is commonly used in temperature or motion sensing applications. This function is useful when the highest value, corresponding to the device closest to the event epicenter, represents the critical or most relevant information, and other values are considered redundant or less important.
Another potential function is a sum with saturation, \textit{i.e.},
\vspace{-2mm}
\begin{equation}
        \displaystyle{{I_{i} = \min \Big(\sum_{j=1}^N I_{i,j}, \Psi\Big).}}
    \vspace{-2mm}
    \label{sum_saturation}
\end{equation}
This function applies to situations where the effect of observations increases to a certain point and then levels off.  
Specifically, scenarios 
such as pollution sensors where the health risks increase as the concentration of pollutants rises, but they level off once a maximum threshold is reached. Similarly, in acoustic sensors, where noise impact on health increases with intensity but eventually saturates. Additionally, in data collection, bandwidth usage increases as more data is transmitted, but it reaches a point of saturation at the network's maximum capacity.  
Here, the piecewise function can cap the total information collected to prevent overloading the BS. 
Finally, controlling the number of active IoTDs may be necessary to conserve energy, particularly in battery-powered devices,  and this approach also helps in managing that. 

\begin{figure}[t]
\centerline{\includegraphics[width=0.8\columnwidth]{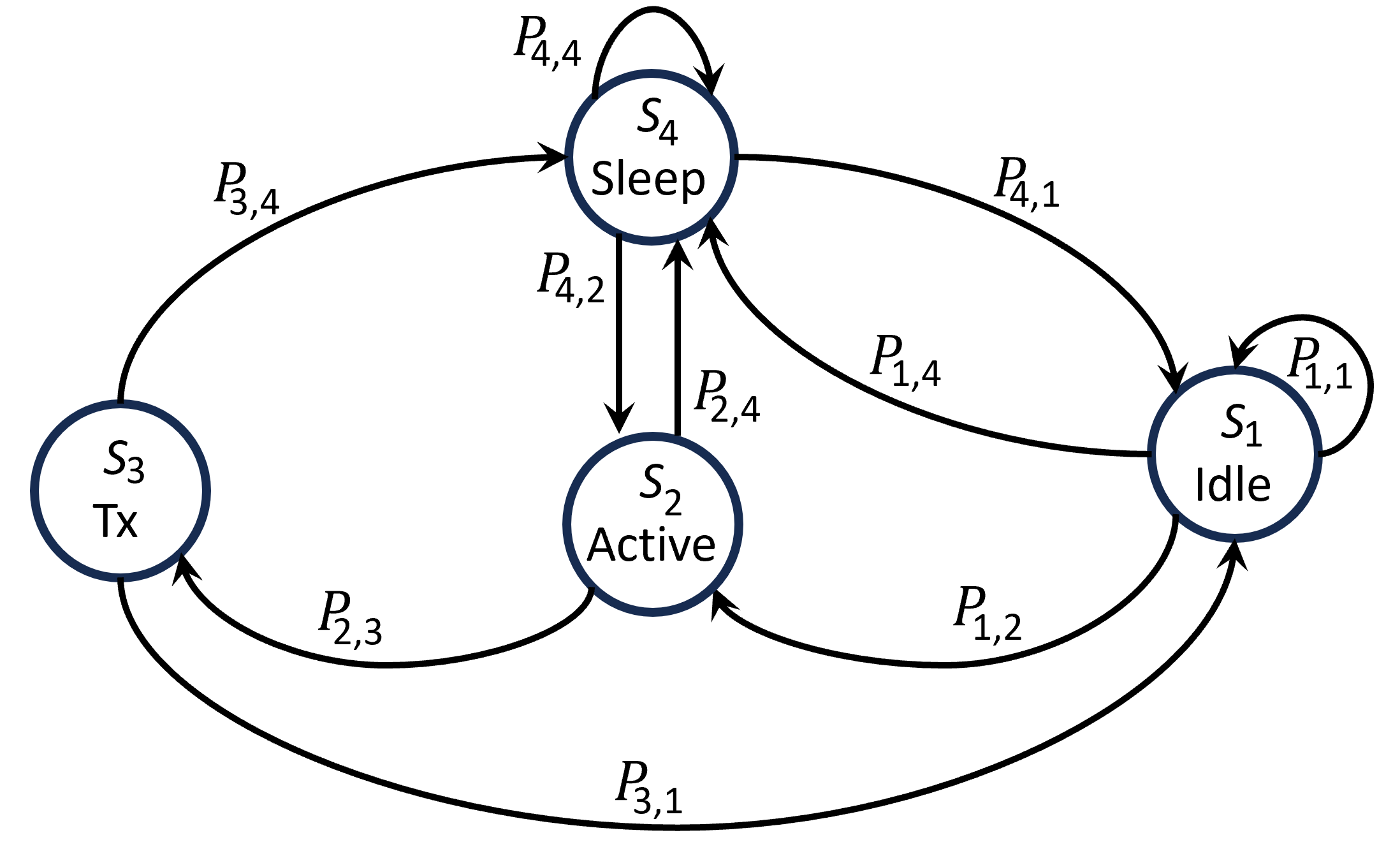}}
\vspace{-3mm}
\caption{The IoTD's operation states are modeled as a four-state discrete-time Markov chain.}
\vspace{-2mm}
\label{fig2}
\end{figure}



\section{Energy Modelling}\label{battery}

This section presents a discrete-time model of battery dynamics powered by ambient EH. We characterize the energy consumption and EH processes, followed by a Markov-based framework to capture battery state evolution and evaluate the probability of sufficient energy availability for operation. 

\subsection{Energy Consumption and Harvesting Model}

We assume discrete EH and storage for device batteries, which have maximum capacity $E_{\text{max}}$. 
Additionally, transmitting in state $S_2$ results in a depletion of $E_\text{Tx}$ units, while monitoring events in state $S_1$ consumes $E_\text{idle}$ units per TTI, and the WuR in state $S_2$ consumes $E_\text{WuR}$, as depicted in Fig.~\ref{fig3}. 
We assume that if the battery level ($B$) of an IoTD at the moment of transmission is lower than the level needed to transmit or perform event monitoring, the battery is depleted and the action is unsuccessful. 

In this paper, we adopt a simplified model based on~\cite{ruiz2022energy}, where the mean energy consumption ($\overline{\mathcal{C}}$) is calculated based on the energy consumption of battery units ($\mathcal{C}_{m}$) for every state $S_{m}$. This is summarized as the mean energy units consumed per IoTD per TTI and written as  
\begin{equation}
    \overline{\mathcal{C}} \!=\! \frac{1}{N}{\sum\limits_{j=1}^{{N}}\! \sum\limits_{m=1}^{4}\! \Pr({S^{(j)}=S_m})\mathcal{C}_{m}}. 
\label{EC}	
\end{equation}

\begin{figure}[t]
\centerline{\includegraphics[width=1.05\columnwidth]{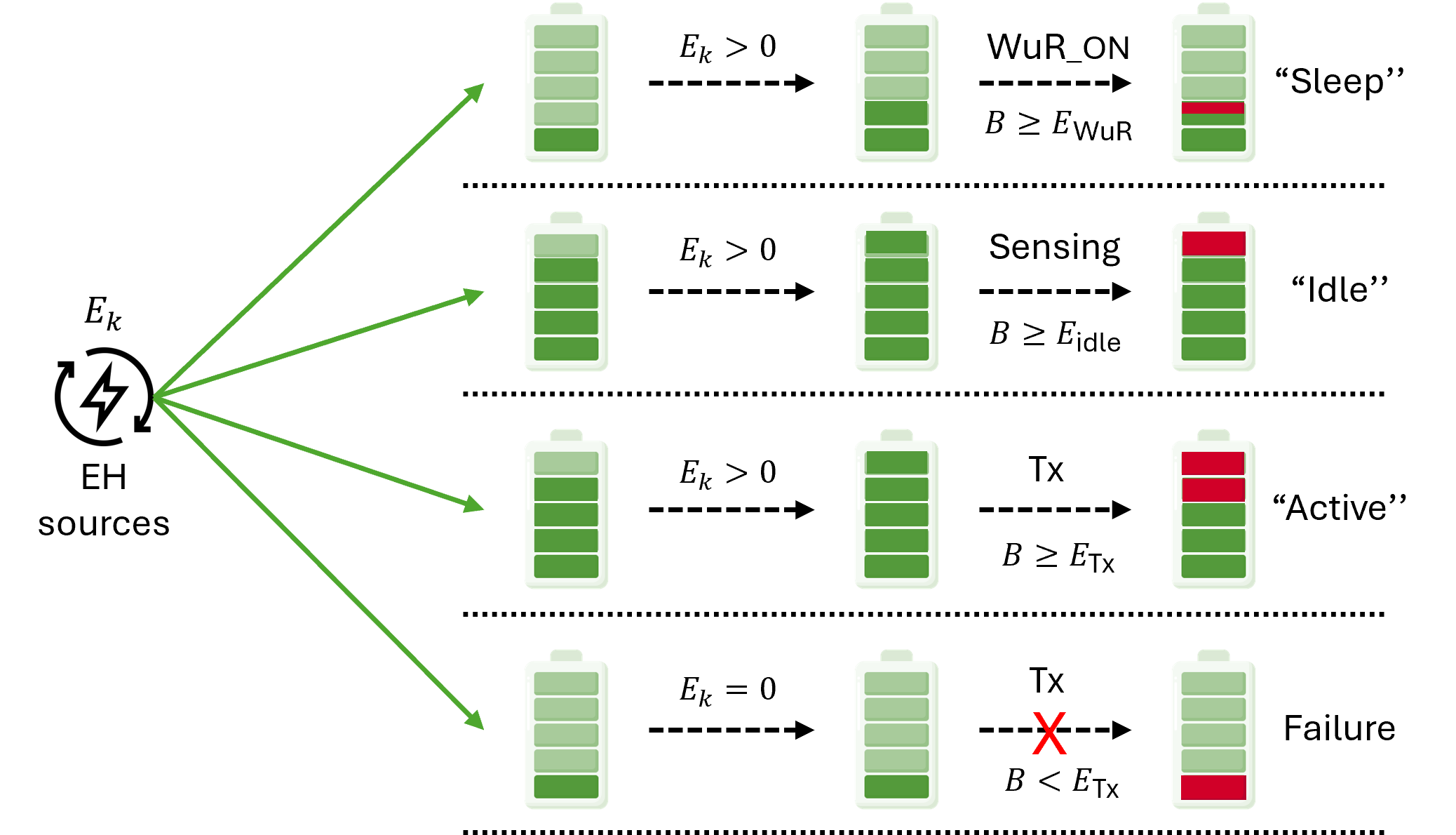}}
\vspace{-3mm}
\caption{{Illustration of the battery level evolution by considering the energy consumption and EH models. Transmission (Tx) results in the highest energy depletion, while ``sleep'' is a low-energy consumption state. The sensing results in a depletion smaller than Tx but greater than a ``sleep'' state. IoTDs that do not have sufficient energy to carry out these actions will fail.}}
\label{fig3}
\vspace{-3mm}
\end{figure}

At each TTI, we also model the energy harvested by the IoTDs by using ${{N}}$ modulated Poisson processes (MPPs), one for each IoTD. 
Specifically, a Poisson process is used to model the occurrence of energy arrivals within a certain time interval. The energy arrival duration, \textit{i.e.}, the time duration in case the energy source remains active, is modeled with an exponential distribution with  mean $\mu_j$. 
At each TTI, at most one energy arrival occurs from an energy source. In that case, the energy source can be considered as active or inactive within one time slot as shown in Fig.~\ref{fig3}. By defining $\lambda_j = 1/\mu_j$, the probability of the energy source being active within one TTI can be expressed as
\vspace{-2mm}
\begin{equation}
    p_{j} = \lambda_j \tau \exp(-\lambda_j \tau). 
    \vspace{-3mm}
    \label{poissonenergy}
\end{equation}

\vspace{-2mm}
\subsection{Battery State Evolution}

The probability that an IoTD has sufficient energy in the battery to perform transmission or sensing depends on the energy harvested at each TTI plus the initial battery level and previous actions the IoTD performs. To calculate this probability, we can use a discrete-time Markov chain to model the battery level over time. We define several states for the battery level, while transitions occur based on the EH process and battery consumption operations. 

Let us define $E_\text{B}$ as the units of energy an IoTD can harvest when the energy source is active. Then, following MPP, harvesting $E_\text{B}$ units of energy occurs with probability $\exp{(-\lambda_j)}$.  
We formulate a discrete Markov chain with states representing the battery level (from 0 to $E_\text{max}$ units), where $E_\text{max}$ is the maximum capacity of the IoTD battery. Let us denote the states of the battery level 
as follows:
\begin{itemize}
    \item State 0: Battery level is 0 units;
    \item State 1: Battery level is $E_\lambda$ units;
    \item State 2: Battery level is $2E_\lambda$ units;
    \item \dots
    \item State $\left\lceil E_\text{max}/E_\lambda \right\rceil$: Battery level is $E_\text{max}$ units; 
\end{itemize}
where $E_\lambda = E_\text{B}\exp{(-\lambda_j)}$. 
We calculate the transition probabilities between these states based on the EH process and battery consumption operations. Assuming a steady state, the state arrival rate equals the state departure rate. Therefore, we can use the fact that $\sum_{b = 0}^{\left\lceil E_\text{max}/E_\lambda \right\rceil} b = 1$, where $b$ is the steady $B$ state probability, to solve $\mathbf{R_{B,B}} \times \mathbf{b} = \mathbf{c}$, where ${\mathbf{c} = [1, 1, \cdots, 1]^\mathrm{T}}$, ${\mathbf{b} = [b_0, b_1, \cdots, b_{\left\lceil E_\text{max}/E_\lambda \right\rceil}]^\mathrm{T}}$, and 
\begin{align}
    \!\!\mathbf{R_{B,B}} \!=\!\! \begin{pmatrix}
{R_{0,0}} & {R_{0,1}} & \cdots & {R_{0,\left\lceil \frac{E_\text{max}}{E_\lambda}
\right\rceil}} \\
{R_{1,0}} & {R_{1,1}} & \cdots & {R_{1,\left\lceil \frac{E_\text{max}}{E_\lambda}
\right\rceil}} \\
\vdots & \vdots & \ddots & \vdots \\
{R_{\left\lceil \frac{E_\text{max}}{E_\lambda}
\right\rceil,0}} & 
{R_{\left\lceil \frac{E_\text{max}}{E_\lambda}
\right\rceil,1}} & \cdots 
& {R_{\left\lceil \frac{E_\text{max}}{E_\lambda}
\right\rceil,\left\lceil \frac{E_\text{max}}{E_\lambda}
\right\rceil}} \\
1 & 1 & \cdots & 1\\
\end{pmatrix}  
    \label{Markov_Battery3}
\end{align}
is the transition matrix. Herein, $R_{k,l}$ represents the transition probability from a battery state $B = k$ to a battery state $B = l$. 

Now, we proceed to find the probability of being in a state with sufficient battery to perform the operation that needs $E_\text{Tx}$ units, \textit{i.e.,} $\Pr(B\ge E_\text{Tx})$. 
Note that IoTD $j$ goes to state $S_2$ with probability $P_{1,2}^{(j)} + P_{4,2}^{(j)}$. While in $S_2$, we need to check whether the IoTD has sufficient energy to transmit the sensed information.  
Exploiting 
the previously presented Markov chain 
for the battery state evolution,
we obtain 
\begin{align}
    \!\!\!\Pr(B^{(j)}\!\!\ge\! E_\text{Tx}) \!=\!\!\!\! &\sum_{B = {E_\text{Tx}}}^{{\left\lceil E_\text{max}/E_\lambda \right\rceil}} \!\!\!\binom{B}{E_\text{Tx}} 
    \frac{\big(P_{1,2}^{(j)} + P_{4,2}^{(j)}\big)^{E_\text{Tx}}}{\big(1 - P_{1,2}^{(j)} - P_{4,2}^{(j)}\big)^{E_\text{Tx}-B}},  
    \vspace{-5mm}
    \label{Markov_Battery}
\end{align}
{where the binomial coefficient represents the number of ways the IoTD can accumulate enough energy to transmit $E_\text{Tx}$, and the transition probabilities account for the activation probabilities that lead or not to consuming $E_\text{Tx}$.} 
This expression has a semi-closed form using the regularized incomplete beta function, $\beta_{1 - P_{1,2}^{(j)} - P_{4,2}^{(j)}}(E_{\text{Tx}} + 1, E_\text{max}+1)$,  with parameters $E_{\text{Tx}} + 1$ and $E_\text{max} + 1$~\cite{egorova2023computation}. 
We can similarly calculate $\Pr(B^{(j)}\ge E_\text{idle})$, \textit{i.e.}, the probability that the IoTD can sense in state $S_1$. 

\section{On-demand Wake-up and Transition Probabilities}\label{analisis}

We assume the IoTDs are equipped with WuR modules to detect requests from the BS to either enter the ``active'' state ($S_2$) or switch between states. The WuR module allows IoTDs with low-energy availability to remain in state $S_4$ while ensuring their availability to provide additional information to the BS when required (with probability $P_{4,2}^{(j)}$).
The BS may try to wake up certain IoTDs in state $S_4$ to sense potential events based on information received from other IoTDs and their spatial correlation. However, this might not always be possible because each IoTD $j$ with low-energy availability stays in state $S_4$ (with probability $P_{4,4}^{(j)}$) and harvests energy for future events. Meanwhile, IoTDs in state $S_4$ with high-energy availability may be authorized by the BS to move from state $S_4$ to $S_1$ to monitor potential events. 
Then, the transitions 
between states for IoTD $j$ can occur as follows. 
\begin{itemize}[leftmargin=8pt, labelsep=2pt, itemsep=1pt]
    \item A transition from state $S_4$ to $S_1$ occurs during the ON time of the duty cycling, $t_{on}$, then 
    \begin{equation}
        P_{4,1}^{(j)} = {\text{$t_{on}$}}/{\text{$t_{DRX}$}},
        \vspace{-0.5mm}
    \end{equation} 
    where $t_{DRX}$ is for the time between consecutive ON states ($S_1$). Otherwise, the IoTD stays in state $S_4$ with probability ${P_{4,4}^{(j)} = 1 - P_{4,1}^{(j)} - P_{4,2}^{(j)}}$. Note that $t_{on}$ and $t_{DRX}$ are integer multiples of $\tau$.
    
    \item A transition from state $S_4$ to $S_2$ occurs when the BS determines that additional information about an event is needed.  
    Based on the current information $I_i$, the learned spatial correlation, and the probability $\Pr(B^{(j)} \!\ge\! E_\text{Tx})$, the BS sends a wake-up request. Upon receiving it via the WuR, the IoTD transitions to state $S_2$, activating both its sensor and main RFI modules.  
    To determine the transition probability, we use the conditional probability $\Pr(S^{(j)}=S_3|S^{(h)}=S_3)$, which represents the probability that an IoTD $j$ is in state $S_3$ given that another IoTD $h$ is in state $S_3$. 
    Then, 
    \begin{equation}
        {P_{4,2}^{(j)} =  \max 
        \Pr(S^{(j)}=S_3|S^{(h)}=S_3)
        , }
    \end{equation}
    wherein the maximum value captures the highest activation probability when an event is detected.  
    Herein, $\Pr(S^{(j)}=S_3|S^{(h)}=S_3)$ is determined by the spatial correlation between IoTD $j$ and each IoTD $h$ in state $S_3$. We calculate $\Pr(S^{(j)}=S_3|S^{(h)}=S_3)$ using the cosine rule as  
\begin{equation}\label{cosine_rule}
   {d_{i,j}^2} = {d_{i,h}^2} + \gamma_{j,h}^2 - 2d_{i,h}\gamma_{j,h}\cos{\varphi},
\end{equation}
where $\gamma_{j,h}$ denotes the distance between both devices and $\varphi$ is the angle between vectors $\vec{d_{i,h}}$ and $\vec{\gamma_{j,h}}$, as depicted in Fig.~\ref{fig1}. Then, 
\begin{equation}\label{cosine_rule2}
   \Pr\!\left(\! S^{(j)}\!=\!S_3|S^{(h)}\!=\!S_3 \!\right)\! \!=\! 
   p{\left(\! \sqrt{{d_{i,h}^2} \!+\! \gamma_{j,h}^2 \!-\! 2{d_{i,h}}\gamma_{j,h}\cos{\varphi}} \right)},
\end{equation}
where $d_{i,h}$ and $\gamma_{j,h}$ are known, and $\varphi$ has a uniform distribution
in $[0, 2\pi)$. 
    
    \item A transition from state $S_{1}$ to $S_{2}$ for an event $(i)$ occurs with probability {${P_{1,2}^{(j)} = \alpha p(d_{i,j})}$}. Otherwise, the IoTD stays in state $S_1$ with probability 
    \begin{equation}
        {P_{1,1}^{(j)} = \sum_{i=\tau}^{\text{$t_{on}$}} (1 - \alpha)^i \frac{\text{$t_{on}$}-i}{\text{$t_{DRX}$}-i} },
    \end{equation}
    while transits from state $S_{1}$ to $S_{4}$ with probabilty ${P_{1,4}^{(j)} = 1 - P_{1,1}^{(j)} - P_{1,2}^{(j)}}$. 
    
    \item A transition from state $S_{2}$ to $S_{3}$ occurs when $B$ is enough to transmit, thus $P_{2,3}^{(j)} = \Pr(B\ge E_\text{Tx})$, while the device transits from state $S_{2}$ to $S_{4}$ with probability $P_{2,4}^{(j)} = 1 - P_{2,3}^{(j)}$. 
    
    \item A transition from state $S_{3}$ to $S_{1}$ occurs with probabilty 
    \begin{equation}
        {P_{3,1}^{(j)} = \sum_{i=\tau}^{\text{$t_{on}$}} \frac{\text{$t_{on}$}-i}{\text{$t_{DRX}$}-i}},
    \end{equation}
    Otherwise, the transition is to state $S_4$ with probability $P_{3,4}^{(j)} = 1 - P_{3,1}^{(j)}$. 
\end{itemize}

\section{Optimization Framework}\label{framework}

An IoTD duty cycling and wake-up management policy controlled by the BS is referred to as ${\boldsymbol{\delta}(k) = [\delta_1(k), \delta_2(k), \dots, \delta_N(k)]}$ for each TTI $k \in [0,K]$, where $K$ represents the total number of TTIs during analysis. At each TTI $k$, 
$\boldsymbol{\delta}(k)$ 
designates which IoTDs will be in $S_1$, waiting to be triggered or woken up from $S_4$ to $S_2$ (designated by 1), and which IoTDs will remain in $S_4$ (designated by 0).   
Here, $\boldsymbol{\delta}(k)$ must take into account ${\Pr(B^{(j)}\ge E_\text{idle})}$ and ${\Pr(B^{(j)}\ge E_\text{Tx})}$ due to the energy arrivals at random times and the finite battery storage capacity. 
This approach helps save power by not prompting an IoTD with low energy levels to wake up and transmit information. Such an action would otherwise drain the battery, preventing the IoTD from transferring the information detected. 


In a long time interval (\textit{i.e.,} large $K$), 
maximizing the information about event $I_{i}$ and minimizing the probability of missing events 
mainly depends on the energy available in the IoTDs batteries. 
Reducing energy consumption means having a greater density of IoTDs with high energy availability for event sensing and reporting. 
However, we need to carefully balance energy savings to ensure high energy availability when needed, while also keeping a sufficient number of IoTDs sensing and reporting to cover a large area and prevent events from going undetected. 
 This, in turn, leads to fewer blind spots where events may go unnoticed, consequently decreasing the misdetection probability. In addition, it increases the probability that the IoTD closest to the event has high energy availability, increasing the received information about the event. 

The proposed optimization problem is stated as
\begin{equation}
        \text{P}1: \text{ } \max_{\boldsymbol{\delta}(k)} \text{ } \frac{1}{\alpha K}\sum_{\forall {i}} I_{i}.  
\label{P1_I}
\end{equation}
Here, we use the expected number of events, $\alpha K$, as a normalization factor instead of the exact number of events, since the latter cannot be known, while $\alpha$ can be inferred from historical data. Note that false alarm reports do not contribute to \eqref{P1_I} 
as they don't lead to a positive $I_{i,j}$ for any $j$. 
In addition, IoTDs with limited energy availability cannot report, so ensuring high energy availability is a crucial condition to maximize information in \eqref{P1_I}. 


\textcolor{black}{Inherently, solving problem P1 in~\eqref{P1_I} entails enhancing battery lifetime within the network. This is because maximizing the information about events necessitates maintaining the devices' battery levels above the threshold required for transmission. Only these devices can transmit event information. Therefore, by maximizing the received information the energy efficiency of the network is also promoted.}

  

{Optimization methods to solve this problem can be computationally intensive and time-consuming. The complexity of solving the problem increases significantly as the number of IoTDs grows~\cite{TCOM_David}. 
Specifically, solving \eqref{P1_I} 
would require a detailed analysis of the system's evolution over time and the IoTDs' energy availability, which is not practical.   
In light of this, we aim to propose heuristic and data-driven approaches to address the issue in a more computationally efficient way. 

\subsection{Heuristic \& KNN-based Duty Cycling}\label{proposal}


In our recent conference work~\cite{ruiz2024intelligent}, we established a minimum expected information level per event, denoted as $I_{\min}$, while aiming to minimize the average energy consumption, denoted as $\overline{\mathcal{C}}$. Our goal here is to solve P1~\eqref{P1_I} by reframing the heuristic approximations based on exhaustive search and KNN presented in~\cite{ruiz2024intelligent}, which is not straightforward. 
The main challenge lies in the limitations of the adjusted duty cycling and KNN approaches when applied to P1~\eqref{P1_I}. Specifically, these methods lack effective mechanisms to control and balance the trade-off necessary to maintain the IoTDs with high energy availability. To tackle the issue, we first determine a realistic target for $\frac{1}{\alpha K}\sum_{\forall {i}} I_{i}$ in~\eqref{P1_I} based on the spatial deployment of the IoTDs. We then calculate the minimum energy required to achieve this information level using the the adjusted duty cycling and
KNN-based heuristic methods.
Our objective is to prevent the depletion of the IoTDs' energy reserves while ensuring that the desired realistic information level is attained following event detection. 


{We formulate the optimization problem in P1 with the framework from~\cite{ruiz2024intelligent} to minimize the energy consumption while achieving the expected information level from the events, $I_{{\min}}$, for each spatial deployment as 
\begin{subequations}
\begin{alignat}{4}
        \text{P}2: \text{ } &\min _{\boldsymbol{\delta}(k)} 
        \text{ } 
        &&\overline{\mathcal{C}}\\
        &\text{ } \text{s.t.} &&\frac{1}{\alpha K}\sum_{\forall {i}} I_{i} \geq  I_{\min}. 
\end{alignat}
\label{P1}
\end{subequations}
\hspace{-2mm}
Herein, 
$\overline{\mathcal{C}}$ can be reformulated from~\eqref{EC} as 
\begin{align}
    \overline{\mathcal{C}} = \frac{1}{NK} \sum\limits_{k=1}^{{K}} 
    \sum\limits_{j=1}^{N} {\delta_j}(k) 
    (E_{\text{idle}} &+ \Pr(S^{(j)}=S_3) E_{\text{Tx}}) + \dots \nonumber\\
    \dots &+ 
    (1-{\delta_j}(k))E_\text{WuR}. 
    \label{EC2}
\end{align}
We assume that state $S_2$ is a conditional transition state so the energy consumption in that state is neglected. 
Note that at each TTI $k$, IoTDs with $\delta_j = 1$ need high energy availability for sensing/transmitting while IoTDs with $\delta_j = 0$ have a low energy consumption in ``sleep'' state $S_4$ related to the WuR consumption, $E_\text{WuR}$. Therefore, proper $\boldsymbol{\delta}(k)$ management is crucial at each TTI.}

\begin{figure}[t]
\centerline{\includegraphics[width=0.6\columnwidth]{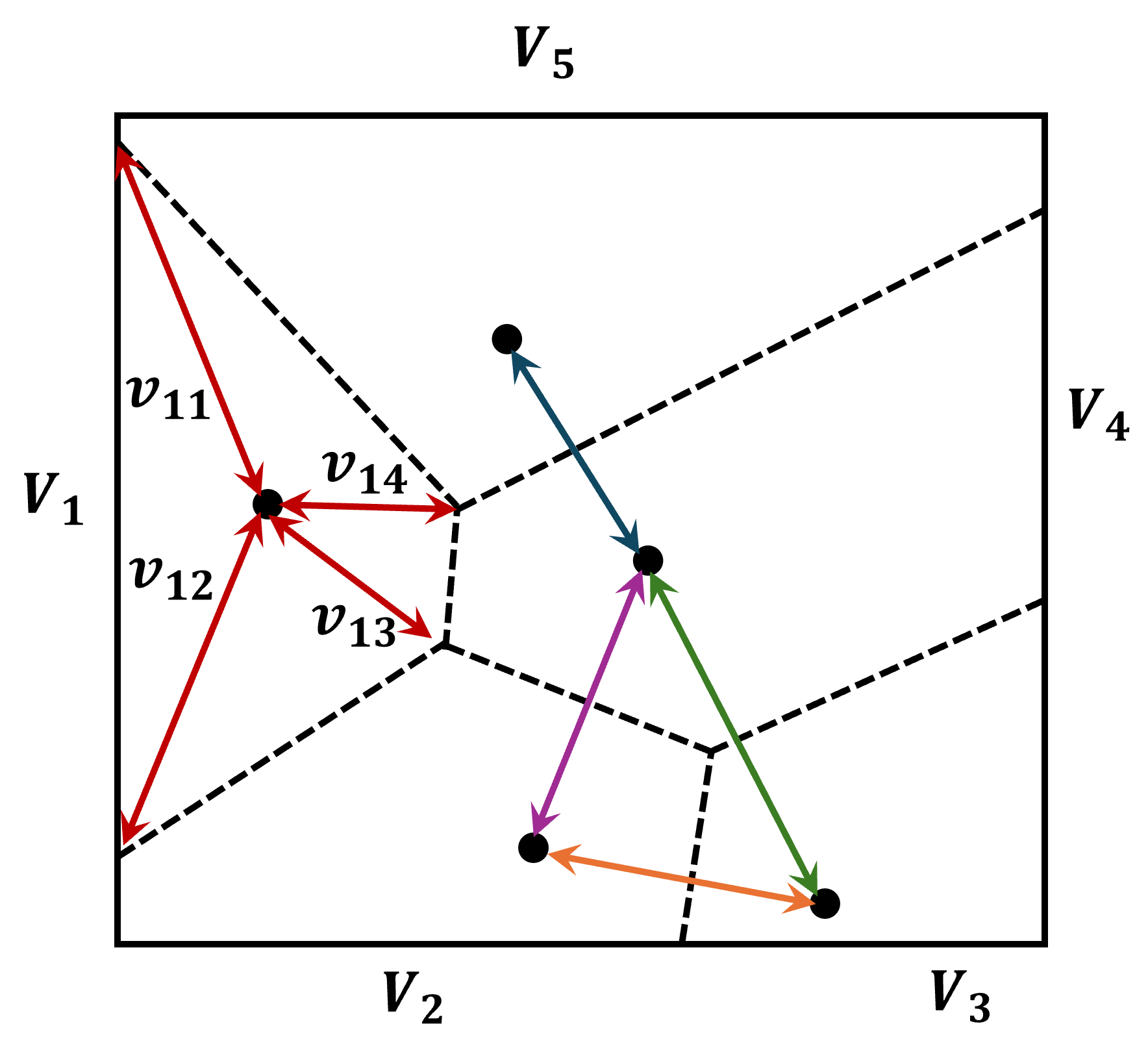}}
\vspace{-3mm}
\caption{{Illustration of a Voronoi diagram for a grid IoT deployment with five devices (represented as black dots). The black dashed lines represent the perpendicular bisectors of the arrows that connect neighboring IoTDs, which form the Voronoi polygon corresponding to each IoTD, $V_j$. The red arrows represent the distances to the vertices, $v_j$, of a Voronoi cell.}}
\label{voronoi_fig}
\vspace{-3mm}
\end{figure}
Here, we calculate the expected value of $I_{{\min}}$ for the network deployment based on the spatial positioning of IoTDs. This contrasts with~\cite{ruiz2024intelligent}, where this value must be given beforehand. 
To calculate $\mathbb{E}(I_{{\min}})$, we construct the Voronoi diagram for the set of $N$ IoTDs positions, as shown in Fig.~\ref{voronoi_fig}. Each cell in the Voronoi diagram contains all points that are closer to a particular IoTD than to any other device, $V_j$. Next, for each IoTD $j$, we find the vertices of its Voronoi cell, $v_j = \{ \mathbf{v}_{j1}, \mathbf{v}_{j2}, \dots, \mathbf{v}_{j\kappa}\}$, where $\kappa$ is the total number of vertices for the Voronoi cell $V_j$. 
These vertices are intersections of the perpendicular bisectors of the lines connecting the device to its neighboring devices~\cite{Voronoi_ma2021escvad}. 
Then, $\mathbb{E}(I_{{\min}})$ is calculated using the maximum of these distances across all Voronoi cells $d^*$ as: 
        $\mathbb{E}(I_{{\min}}) = p\left(d^*\right)$, 
where 
\begin{align}\label{E_Info3}
        d^* = 
        \max_{j} \sqrt{(x_j - x_{v_{j}})^2 + (y_j - y_{v_{j}})^2}.
\end{align}
A particular example is the deployment of an IoTD grid, for which 
we have 
\begin{align}\label{E_Info1}
        \mathbb{E}(I_{{\min}}) = p\left(\sqrt{{\xi}/{(2N)}}\right)
\end{align}
Here, the sensing area assigned to each IoTD is a square area of side $\sqrt{\xi/N}$ and the maximum distance from the center (\textit{i.e.}, IoTD) to the side of this area is $\sqrt{{\xi}/{(2N)}}$. 

Next, we outline two potential approaches to tackle the current problem using the previously calculated $\mathbb{E}(I_{{\min}})$.  
\begin{enumerate}
    \item We adjust the IoTD duty cycling by an exhaustive search within $t_{on}$ ($S_1$) and $t_{DRX}$ values. The duration of $t_{on}$ ($S_1$) plus ``sleep'' state ($S_4$) within a cycle constitutes the $t_{DRX}$ value. Hereinafter, we refer to this approach as ``adjusted duty cycling''. 
    \item We form clusters based on each device's spatial placement. We  create a graph structure using KNN method as follows: {{1) set the assignment of IoTDs to each cluster randomly, 2) calculate the distance to the neighbors, 3) identify each IoTD KNN based on the distances, and 4) assign each IoTD to the cluster that the majority of its KNN belong to, and 5) update centroids and repeat steps 2-5 until convergence to achieve the correct IoTD assignment to each cluster~\cite{Voronoi_ma2021escvad}.}} The algorithm sets a maximum distance $d_{\text{max}}$, linked to the minimum sensing power $p(d_{\text{max}})$ set to form the clusters. The aim is to reduce the misdetection probability by limiting the sensing area controlled by the clusters. Then, we configure the duty cycling of the IoTDs in each cluster. In each TTI, the IoTDs within a cluster iterate a round-robin sensing in $S_1$. That is, during each TTI, only one IoTD is actively sensing in $S_1$ while the other IoTDs in the cluster are in $S_4$, ensuring there is always one IoTD in $S_1$ during each TTI. In each cluster, the $t_{DRX}$ duration for IoTDs within that cluster is equal to the number of IoTDs in that cluster. This proposed approach is summarized in Algorithm~\ref{alg1}. 
\end{enumerate}

Notice that the first approach differs from the second one in the duty cycling configuration. Indeed, the BS wake-up request for the first approach is based on the IoTD's correlation rather than cluster-based, as in Algorithm~\ref{alg1}. 
In both approaches, the BS may request more information about an event from the other IoTDs in state $S_4$ that have a strong spatial correlation with the IoTDs in state $S_3$, line 7. In this way, the BS requests only the needed information using the WuR technology.

\textcolor{black}{Satisfying a minimum expected information per event requires maintaining IoTDs' sensing/reporting availability. That is, maintaining the battery level of IoTDs above the threshold required for transmission.  Only these IoTDs can transmit event information. At the same time, proper sensing cycling and wake-up strategy help to reduce energy consumption and avoid misdetections, which worsen the mean information captured per event.} 
\textcolor{black}{Herein, we ensure that the model maintains the minimum level of information that the BS requires from an event above a threshold ($I_{\min}$) according to the spatial deployment of the IoTDs. Having access to a certain level of information from IoTDs enhances the BS's ability to effectively manage the network, respond to events, and ensure the overall reliability and performance of the IoT system. 
Note that a shorter $t_{DRX}$ decreases the misdetection probability when there is high energy availability; however, it increases the energy consumption and the risk of low energy availability. Then, balancing energy saving and sensing availability to lower the misdetection probability is paramount.}

\begin{algorithm}[t]
\caption{KNN-based solution}\label{alg1}
\small
{\begin{algorithmic}[1]
\State \textbf{Set} number of clusters ($M$) as $\xi/(\pi d_{\text{max}}^2)$
\State \textbf{Form} clusters using KNN method 
\State \textbf{Configure} duty cycling of IoTDs according to its cluster association ($t_{on}$ = 1, duty cycle = cluster size)
    \For{each $k$} 
        \If{BS receives information from ``active'' IoTDs}  
        \State $\displaystyle{\hat{I}_{i} = \max_{h: S^{(h)}=S_3}(I_{i,h})}$, 
            \While{$\hat{I}_{i} < I_{\min}$}
                \State {BS sequentially activates IoTDs ${\forall j: S^{(j)}\!=\!S_4}$} 
                \Statex {\hspace{15.8mm}with ${\Pr\left( S^{(j)}\!=\!S_3|S^{(h)}\!=\!S_3 \right) \geq p(d_{\text{max}})}$,} \Statex {\hspace{15.8mm}${\forall h: S^{(h)}\!=\!S_3}$, one at a time in each loop,} \Statex  {\hspace{15.8mm}starting with the one with the highest value,} \Statex {\hspace{15.8mm}while updating $\displaystyle{\hat{I}_{i} = \max_{j: S^{(j)}=S_3}(\hat{I}_{i},I_{i,j})}$}
            \EndWhile 
        \State \textbf{Update} $I_{i} \leftarrow \hat{I}_{i}$
        \EndIf
    \EndFor
\end{algorithmic}}
\end{algorithm}




\subsection{Reinforcement Learning-based Duty Cycling and Wake-up}

Model-free RL is a method for solving decision-making problems and finding optimal solutions in dynamic environments~\cite{RL_1}. Herein, we treat our optimization problem \eqref{P1_I} as an RL challenge, where a conventional $Q$-learning algorithm can be adapted to
learn the optimal policy~\cite{RL_lu2023deep}. Specifically, we consider the IoTD deployment as the environment, while the central controller at the BS functions as the agent in a centralized learning framework. The key components of our RL framework are outlined as follows.

\textit{1)} The focus is on configuring duty cycling to save energy on the devices while preventing misdetections due to events not being detected or events being detected by devices without enough energy. Specifically, let $\mathcal{T}$ denote the system state space. 
The system state at TTI $k$, $\mathcal{T}(k)$, encompasses the status of each device (``sleep'' $S_3$ or ``idle'' $S_2$), the received sensing power from the event denoted by the sensing function $p(d_{i,j})$, the position of the IoTDs, and battery level. Additionally, the positions of the devices and their current thresholds are known. The state of each IoTD at TTI $k$ is defined as
\begin{equation}
    \mathcal{T}(k) =\{ \{\delta_j(k)\}, \{p(d_{i,j})\!\}, \{(x_j,y_j)\}, \{B^{(j)}\} \}_{j\in \mathcal{N}},
    \label{states_RL}
\end{equation}
where $\{p(d_{i,j})\}$ for those devices that did not report is estimated using $P_{4,2}^{(j)}$ according to \eqref{cosine_rule} and \eqref{cosine_rule2} from the spatial correlation with the IoTDs that did report.   
Given an observed state `$\mathcal{T}$', the BS adjusts $\boldsymbol{\delta}(k)$ for each device based on the reward/penalty function. The action space is thus binary, $\delta_j \in [0,1];$ $\forall j \in {\mathcal{N}}$.

In RL, the reward function evaluates the effectiveness of the duty-cycling policy when the BS takes action in the current state. At each learning step, the system performance aligns with the desired objective~\cite{RL_1}. Hence, a reward function capable of enhancing energy efficiency without increasing misdetection is crucial. 
Therefore, the reward function is formulated as
\begin{align}\label{rl_r} 
        r(k) =& \left(1-\frac{\boldsymbol{\delta}(k) \boldsymbol{\mathbf{1}}^\mathrm{T}}{{N}} \right) - \alpha \mu_1 \mathbf{1}_0(I_{i}),
\end{align}
where the coefficient $\mu_1$ is a positive constant for balancing utility and cost. Additionally, the first term represents the energy efficiency factor, and the second one is the misdetection factor.  
Note that the indicator function in the second term imposes the error probability satisfaction level: if no misdetection occurs, then no punishment of the reward function due to any error.

\textit{2)} We focus on the case that the BS might require further information about the event once is detected.
Once an event is detected, the current system state $\mathcal{T}(k)$ encompasses the status of each device (``sleep'' $S_3$ or ``active'' $S_1$), their received sensing power from the event denoted by the sensing function $p(d_{i,j})$, and information on IoTDs' battery level. Additionally, the current state of each device is also defined by \eqref{states_RL}. Given an observed state `$\mathcal{T}$', the BS adjusts $\boldsymbol{\delta^*}(k)$ for each device in case it needs more information about the event. The action space is thus also binary, $\delta^*_j \in [0,1];$ $\forall j \in {\mathcal{N}}$.
At this stage, the reward function is formulated as
\begin{align} \label{rl_r*}
        r^*(k) =& \mu_2 I_{i} - {\boldsymbol{\delta^*}(k) {\mathbf{1}}^{\mathrm{T}}}/{{N}},
\end{align}
where $\mu_2$ is a positive constant for balancing utility and cost, imposing a satisfaction level for $I_{i}$ and punishing low information flow. Meanwhile, the second term accounts for the efficiency in waking up IoTDs previously in $S_3$.  

The goal is to determine an optimal policy $\pi^*$, with ${\pi^*(\mathcal{T}(k)): \mathcal{T}(k) + \pi^* \rightarrow \mathcal{T}(k+1)})$~\cite{RL_lu2023deep}, that maximizes the long-term expected discounted reward. 
The latter is given by 
\begin{equation}
    \begin{array}{rl}
        U_{k} = \sum_{i = 0}^{k} \zeta^i r_{i+1}, 
    \end{array}
\end{equation}     
where $\zeta \in (0,1]$ is the discount factor whose impact decreases exponentially with each state $k$ and $r_{k+1}$ is the reward at the next state. 
Note that the BS determines $\boldsymbol{\delta}(k)$ whenever an action is required, ensuring that IoTDs are free from any computational burden associated with the decision-making process beyond their predefined duty cycle. The decision-making process is handled entirely by the central BS. Furthermore, this modification is binary, resulting in negligible overhead on the control channel from the BS to the IoTDs.

{\subsection{Decision Transformer-based Duty Cycling and Wake-up}}

\begin{algorithm}[t]
\caption{DT-based solution}\label{algDT}
\small
{\begin{algorithmic}[1]
\State \textbf{Set} initial parameters for the linear transformation model $\mathcal{W}(\cdot)$ and the weight for each episode $\phi(k)_i$
\State \textbf{Set} RL policy \eqref{delta(k)}
    \For{each episode} 
        \State \textbf{Compute} attention mechanism projections 
        \Statex \hspace{5.2mm}(key $\phi(k)_i^{\mathcal{W}_1}$, query $\phi(k)_i^{\mathcal{W}_2}$, and  value $\phi(k)_i^{\mathcal{W}_3}$), and
        \Statex \hspace{5.2mm}\textbf{calculate} attention scores weights using \eqref{delta(k)3}
    \EndFor
\State \textbf{Calculate} and \textbf{concatenate} outputs 
using \eqref{delta(k)2} 
\State \textbf{Update} state $\{\mathcal{T}(k)\}$, and the action $\{a_j(k)\}$, \textit{i.e.,} set new values for $\boldsymbol{\delta}(k+1)$
\State \textbf{Calculate} $\{r(k+1)\}$ and \textbf{update} parameters for DT model, $\mathcal{W}(\cdot)$ and $\phi(k)_i$    
\end{algorithmic}}
\end{algorithm}

RL approaches require significant interaction with the environment for dynamic data collection. This is particularly challenging in domains where interactions are costly, and thus training data acquisition might be limited. As a result, off-line RL has been gaining attention as a data-driven learning method that can derive policies from static datasets collected previously without interaction with the environment~\cite{hu2024transforming}. The offline RL setting can fully 
extract the optimal policy from a large amount of offline data but also needs to address the distribution discrepancy between the offline training data and the target policy. 
However, this makes it harder for agents since they cannot explore the environment or gather more feedback. In this sense, due to the sequential decision process of RL, a natural idea is to use transformers~\cite{vaswani2017attention} as alternative architectures 
to improve RL methods~\cite{hu2024transforming}. 
DT considers the RL problem as a task for modeling conditional sequences and avoids traditional RL challenges, and it is adopted here. Specifically, DT is used to obtain an RL model offline and then use it to predict a decision policy online.   

During DT's training process, we repeatedly sample a batch of tasks to gather experience and thus learn. For each task, we execute a sequence of $Z$ episodes (TTIs). During optimization, we combine these episodes to optimize a single policy model and aim to maximize the discounted cumulative reward of this policy. Specifically, transformers consist of stacked self-attention layers with residual connections. Each layer of self-attention receives $N$ embeddings $\phi(k)$, corresponding to unique input tokens with ${\phi(k) = \{\{\mathcal{T}(k)\}, \{a_{j}(k)\}_{j \in \mathcal{N}}, \{r(k)\}\}}$ representing a working memory at TTI $k$, according to the state $\mathcal{T}(k)$, the action $a_{j}(k) \sim \pi(\mathcal{T}(k))$, and the reward $r(k)$ in~\eqref{rl_r}, as shown in Fig.~\ref{task_fig}a. 
We also have $N$ outputs $\{\delta_j(k)\}_{j \in \mathcal{N}}$, preserving the input dimensions. 

\begin{figure}[t]
\centering
\includegraphics[width=0.9\columnwidth]{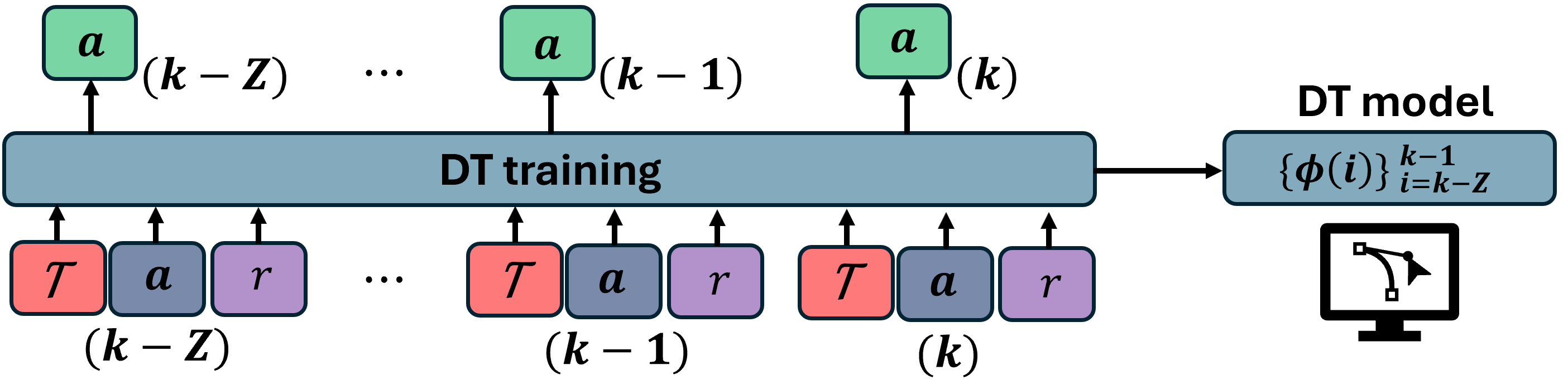}

\includegraphics[width=0.9\columnwidth]{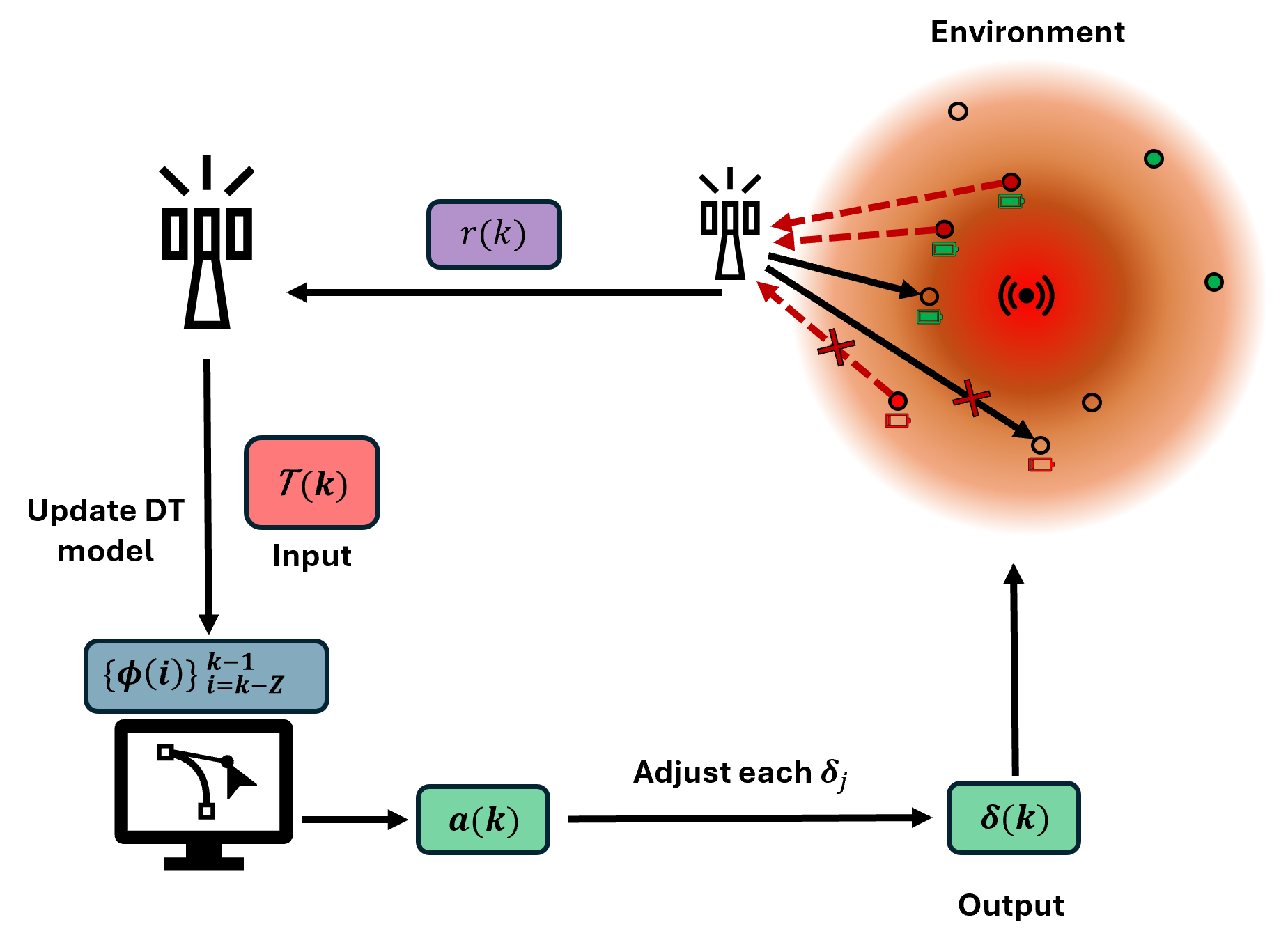}
\vspace{-4mm}
\caption{{Illustration of an RL-based Transformer. a) Learning process for one task of $Z$ episodes using DT-architecture (top), and b) IoT network setup using DT-model (bottom). Note that the DT model is run at the BS.}}
\label{task_fig}
\vspace{-3mm}
\end{figure}
Then, we represent each task as a linear combination of the $Z$ working episodes, $[\phi(k-Z), \phi(k)]$,  sampled as
\begin{align} \label{delta(k)}
        \boldsymbol{\delta}(k) = \sum_{i=1}^Z \psi_i \mathcal{W}(\phi(k)_i), \hspace{2mm}\text{with } \sum_{i=1}^Z \psi_i = 1, 
\end{align}
where $\mathcal{W}(\cdot)$ represents an arbitrary linear transformation. Moreover, $\psi_i$ is a coefficient to indicate how relevant a particular episode is to the task representation given the set of sampled TTIs. 
The input is assigned via linear transformations with $\mathcal{W}(\cdot)$ to a key $\phi(k)_i^{\mathcal{W}_1}$, a query $\phi(k)_i^{\mathcal{W}_2}$, and a value $\phi(k)_i^{\mathcal{W}_3}$. The output of the self-attention layer is then given by weighting the values $\phi(k)_i^{\mathcal{W}_3}$ by the normalized dot product between the query $\phi(k)_i^{\mathcal{W}_2}$ and other keys $\phi(k)_i^{\mathcal{W}_1}$ using the softmax function~\cite{zhu2020efficient}, obtaining  
\vspace{-1mm}
\begin{align} \label{delta(k)2}
        \boldsymbol{\delta}(k) = \sum_{i=1}^Z 
        \text{softmax}\left({ \big{\langle} \phi(k)^{\mathcal{W}_2}_i, \{ \phi(k)^{\mathcal{W}_1}_{i^*}\} \big{\rangle}}
        \right)
        \phi(k)^{\mathcal{W}_3}_i, 
\end{align}
where 
\vspace{-1mm}
\begin{align} \label{delta(k)3}
\text{softmax}\left({ \big{\langle} a, \{ b_{i^*}\} \big{\rangle}}\right)
= \frac{\exp{\left(a\right)}}{\sum_{i^*=1}^Z \exp{\left(b_{i^*}\right)}}.  
\end{align}
This approach allows us to attribute ``credit'' to the layer by creating connections between the state and return through the similarity of the query and key vectors (maximizing the dot product). This enables the model to generate a policy $\boldsymbol{\delta}(k)$ based on the current state $\mathcal{T}(k)$ and the preceding $Z$ episodes $\{\phi(k)_i\}_{i=k-Z}^{k-1}$, as shown in Fig.~\ref{task_fig}b. The proposed approach is summarized
in Algorithm~\ref{algDT}. 

\vspace{5pt}
\subsection{Complexity and Implementation Aspects}

The adjusted duty cycling involves an exhaustive search within $t_{on}$ and $t_{DRX}$. Its computational cost is $\mathcal{O}(\varrho^N)$ where $\varrho$ is the number of possible $t_{on}$ values multiplied by the number of possible $t_{DRX}$ values. On the other hand, the complexity of the approach implementing KNN is bounded to $\mathcal{O}({N}({N}-1)/2)$. Such a worst-case complexity arises due to the distance computation to the ${N}$ IoTD for each query point using a brute-force approach. 
However, as previously mentioned, the computation time can differ based on the utilized algorithm, occasionally reducing to $\mathcal{O}(\min \{M({N} - M), ({N} - M)^2\})$, where $M$ is the number of clusters~\cite{Voronoi_ma2021escvad}. 
Meanwhile, the complexity of the RL-based approach might vary depending on the initial values and the algorithm implementation. However, the worst-case complexity is  $\mathcal{O}(N^2)$. Similarly, the computational complexity of the self-attention mechanism in transformers increases quadratically with the input sequence length ($N$). 
The exact complexity depends on the specific algorithms and how fast they converge. 
Notice that RL is the only one that requires online implementation or a quite large dataset at least. 

Note that the policy $\boldsymbol{\delta}(k)$ does not require updates at every TTI. Once the approaches are trained and/or implemented, the sensing policy remains fixed for each specific spatial deployment. Updates from the BS to the IoTDs are only necessary when an event is detected or periodically for synchronization purposes. However, these synchronization updates occur at intervals significantly longer than typical TTI durations. Furthermore, as mentioned previously, these updates are binary, resulting in negligible overhead on the control channel between the BS and the IoTDs.
Moreover, the learning process for each approach represents a one-time energy cost at the BS, while the resulting duty cycling and EH operations ensure sustained long-term energy efficiency at the IoTDs. 





\vspace{-1ex}
\section{Simulations}\label{result}

\begin{table}[t!]
\caption{Simulation parameters}
\label{table_sim}
\begingroup \centering
    \begin{tabular}{lll}
    \hline
    \textbf{Parameter}                       & \textbf{Value}     &\textbf{Ref.}\\
    \hline
    $d_{max}$& 4 meters&\cite{ul2022learning}\\
    $\eta$&1&\cite{thomsen2017traffic, ruiz2022energy}\\
    $\Psi$&1&\cite{thomsen2017traffic, ruiz2022energy}\\
    $\xi$&$20\times20$&\cite{thomsen2017traffic,ruiz2022energy}$^*$\\
    $E_{\text{max}}$&100 units&\cite{ruiz2022energy,howitt2005extended}$^\dag$\\
    $E_{\text{idle}}$&1 units&\cite{rostami2020novel,ruiz2022energy,howitt2005extended}$^\dag$\\
    $E_{\text{WuR}}$&1/14 units&\cite{rostami2020novel,ruiz2022energy}$^\dag$\\
    $E_{\text{Tx}}$&10 units&\cite{rostami2020novel,ruiz2022energy,howitt2005extended}$^\dag$\\
    $p(d_{i,j})$&$e^{-\eta d_{i,j}}$&\cite{thomsen2017traffic, ruiz2022energy}\\
    $N$&[10, 250]&\cite{thomsen2017traffic,ruiz2022energy}$^*$\\ 
    \hline
    \end{tabular}{}\\
    \endgroup
    \vspace{1ex} 
    \footnotesize{$^*${The values of $\xi$ and $N$ lead to a density range [0.025, 0.625] IoTDs/$m^2$.}}\\
    \footnotesize{$^\dag${Energy parameters are normalized from typical values.}}
\vspace{-2ex}
\end{table}
  
We consider 
a 20$\times$20 m$^2$ area with ${N} \in 
[10, 250]$. Note that  ${N} < 10$ is not enough to cover the area, leading to many misdetection events, thus we ignore such setups. 
We assume an exponentially decreasing function, i.e., $p(d_{i,j}) = e^{-\eta d_{i,j}}$, with $\eta= 1$~\cite{ruiz2022energy}, and $\Psi = 1$, related to the ideal case when  $d_{i,j} = 0$. 
We perform $10^3$ runs of Monte Carlo simulations, each having different deployments and lasting $10^4$ TTIs. In addition, we assume $E_\text{Tx} = 10$ units while $E_\text{idle} = 1$ unit, $E_\text{WuR}= 1/14$ units, $E_\text{max}= 100$ units,  and $p(d_{\text{max}})=0.018$ corresponding to $d_{\text{max}} = 4$ meters~\cite{ul2022learning}. We assume IoTDs performing EH with 
$\lambda_j = \lambda, \forall j \in \mathcal{N}$. 
Table~\ref{table_sim} summarizes the parameters used in simulations unless explicitly stated otherwise.


We consider two benchmarks. In the first one, we let each device adopt a random duty cycling with $t_{on}$ $\in [1,2]$ and $t_{DRX} \in [2, 4, 8]$ for light sleep and avoiding misdetection~\cite{fowler2012analysis}. 
For the second one, we adopt a genie-aided approach wherein the closest IoTD to the event epicenter always detects the event. If the closest IoTD is more than $d_{\text{max}}$ away, then it is considered a misdetection. It is important to mention that this approach is not feasible in practice because the BS would need to know a priori which IoTD is the closest to the event epicenter to activate it using WuR. Nevertheless, this approach provides an upper bound in performance. We also evaluate the effect of the reward weights $\mu_1 \in [0.5, 20]$ and $\mu_2 \in [1, 5]$ in the RL-based approach. Specifically, $\mu_1$ controls the trade-off between detection accuracy and energy consumption, while $\mu_2$ adjusts the balance between information gain and its associated energy cost. Both weights were optimized using Bayesian optimization~\cite{garnett2023bayesian}. Preliminary results suggest that moderate values within these ranges yield a good balance between reliability and energy efficiency. 



\subsection{Performance Comparison}
 
In Fig.~\ref{Info_fig}, we show the average information received by the BS per event, which increases with the density of IoTDs. This is because more IoTDs are active at the same time, providing more information to the BS. Herein, the proposed methods consistently outperform the random duty cycling benchmark, regardless of the IoTDs density. Specifically, the adjusted duty cycling and the KNN-based proposals can provide up to $\sim$0.4 units of information per event while the RL-based solution approaches the performance of the genie-aided benchmark for high-density scenarios with $\sim$0.72 units of information per event. Note that with the genie-aided method, the BS always obtains the greatest amount of information from the event. 
Similarly, the DT-based proposal outperforms the other methods and the random duty cycling benchmark. However, its performance still falls significantly short compared to the RL-based solution and the genie-aided benchmark. 
On the other hand, for low-density scenarios $N\leq 50$, all proposed methods have similar performance, outperforming the random duty cycling benchmark but without approaching the optimal values given by the genie-aided benchmark.  

\begin{figure}[t]
	\centering
	\includegraphics[width=0.9\columnwidth]{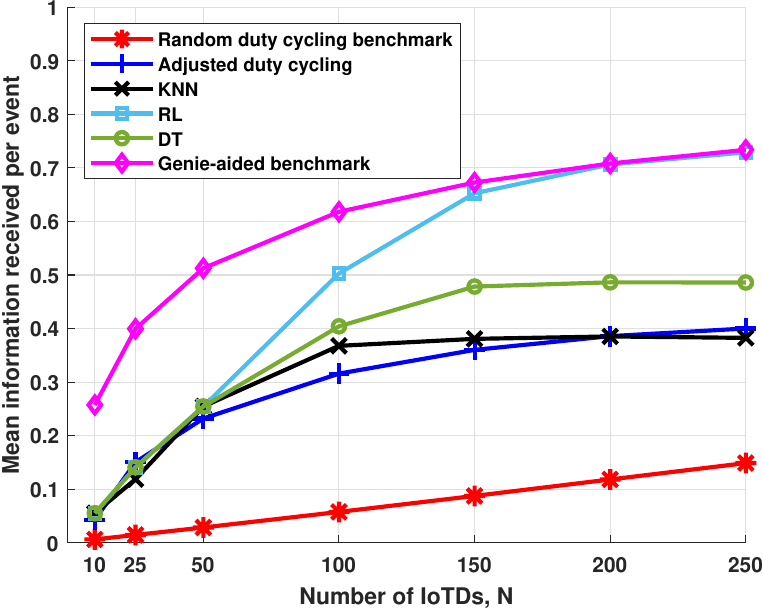}
  \vspace{-3mm}
    \caption{Mean information received by the BS per event as a function of the number of IoTDs.}
	\label{Info_fig}
  \vspace{-3mm}
\end{figure}
 
\begin{figure}[t]
	\centering
	\includegraphics[width=0.9\columnwidth]{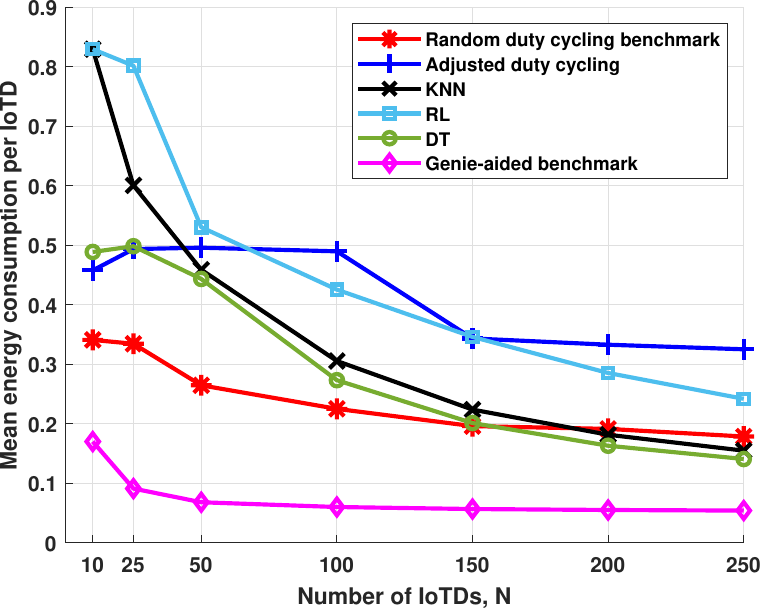}
  \vspace{-3mm}
    \caption{Mean energy consumption per IoTD per TTI as a function of the number of IoTDs.}
	\label{EC_fig}
 \vspace{-3mm}
\end{figure}
\begin{figure}[t]
	\centering
	\includegraphics[width=0.9\columnwidth]{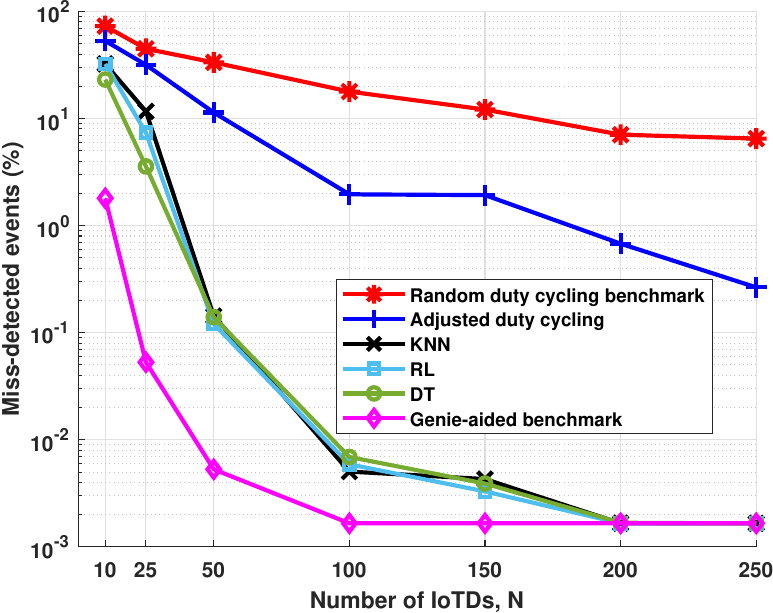}
 \vspace{-3mm}
    \caption{Percentage of misdetections as a function of the number of IoTDs.}
	\label{miss_fig}
 \vspace{-3mm}
\end{figure}
Hereinafter, we also investigate the mean energy consumption per device per TTI as the density of IoTDs increases. The results depicted in Fig.~\ref{EC_fig} show that for a setup with 10 IoTDs the mean energy consumption for the proposals is around 0.45-0.83 units per IoTD per TTI and only 0.34 units for the random duty cycling benchmark due to its poor event-detection performance.  
As the device density increases, the mean energy  consumption decreases up to 0.145 units for the proposed KNN-based method. In contrast, the mean energy consumption for the random duty cycling benchmark is above 0.18 units per TTI, which means that energy consumption increases by more than 29\%.  
Meanwhile, the proposed RL-based solution and the adjusted duty cycling decrease the energy consumption down to 0.24 and 0.32 units, respectively.  
Notably, for more than 180 IoTDs, the mean energy consumption of the proposed KNN-based method is lower than for the random duty cycling benchmark. In addition, Fig.~\ref{EC_fig} shows how the mean energy consumption per device per TTI decreases 40\% and 10\% when using the  DT-based proposal compared to the initial RL-based and the KNN-based methods, respectively. As the deployment density increases, the proposed methods reduce the gap in energy consumption regarding the genie-aided benchmark. In high density scenarios, this can result in a reduction of up to 0.145 units per IoTD per TTI, compared to 0.055 units per IoTD with the genie-aided approach.

 
In Fig.~\ref{miss_fig}, we illustrate the probability of misdetection for each configuration as a function of the IoTDs deployment density. 
As expected, the probability of misdetection decreases with the density. For instance, the misdetection probability is around 33\%, 58\%, and 77\% for the KNN-based approach, adjusted duty cycling approach, and random duty cycling benchmark, respectively, when $N=10$, thus evincing many/large blind spots. However, as the IoTD density increases, the probability drops below 0.01\% for the proposed KNN-based and RL-based solutions. For the random duty cycling benchmark, the misdetection probability drops to a minimum of 7\% for deployments with 250 IoTDs. 
Notice that the KNN-based, RL-based, and DT-based solutions outperform both the adjusted duty-cycle proposal and the random duty cycling benchmark. Even though the adjusted duty cycling can outperform the benchmark by up to a 200\%, just optimizing the duty cycling is not enough to lower the misdetection probability below 2\% for deployments with less than 200 IoTDs. Note that with the genie-aided approach, the probability of misdetection drops from 2\% to near 0.003\% as the IoTD density increases. This is because IoTDs have higher energy availability, as they do not waste energy on sensing while there is no event happening. However, there is still a small probability that the IoTD that detected the event does not have enough energy to transmit. This value, which is $\sim$0.001\%, is independent of the number of IoTDs and constitutes a lower bound.    

\begin{figure}[t]
	\centering
	\includegraphics[width=0.9\columnwidth]{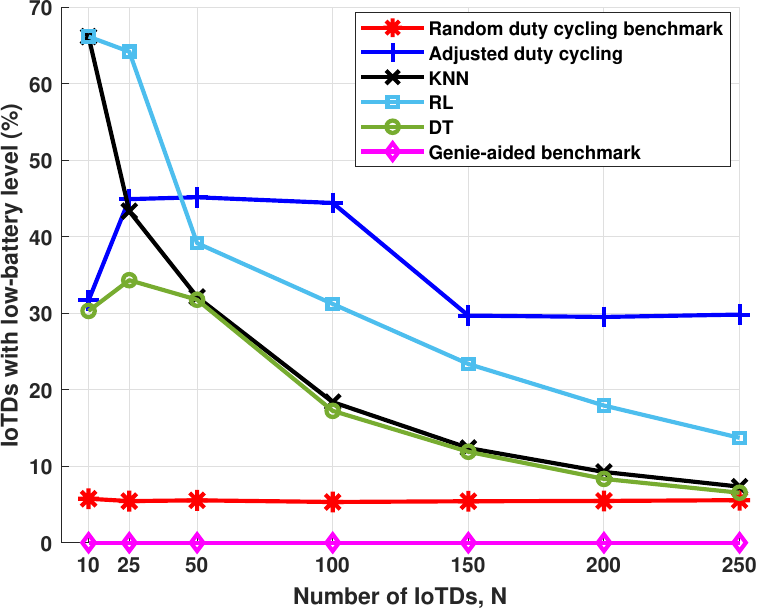}
  \vspace{-3mm}
    \caption{Percentage of IoTDs with low-energy availability to Tx ($B<E_\text{Tx}$) per TTI as a function of the number of IoTDs.}
	\label{NoBat_fig}
  \vspace{-3mm}
\end{figure}
\begin{figure}[t]
	\centering
	\includegraphics[width=0.9\columnwidth]{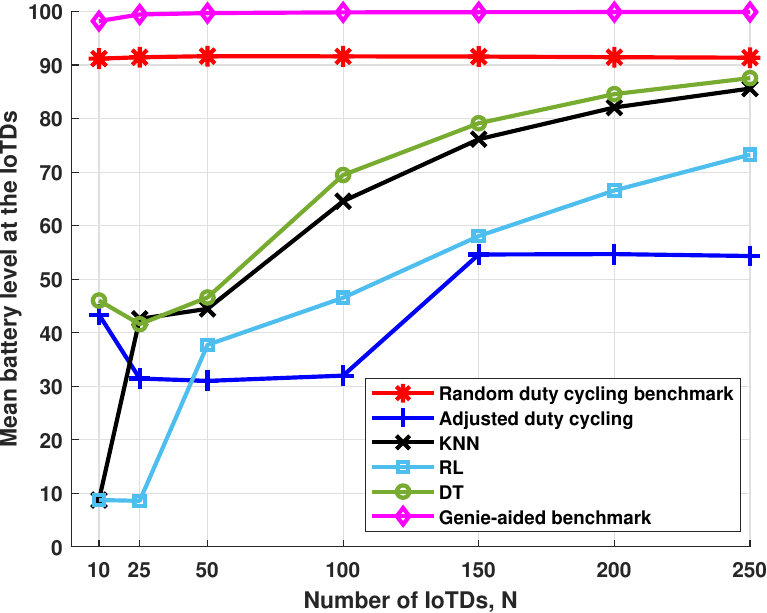}
  \vspace{-3mm}
    \caption{Mean battery level per TTI as a function of the number of IoTDs.}
	\label{Bat_fig}
  \vspace{-3mm}
\end{figure}
Figs.~\ref{NoBat_fig} and \ref{Bat_fig} show the mean number of IoTDs with low-energy availability per TTI and the mean battery level of the IoTDs per TTI. These capture the IoTD availability for sensing and transmission. These performance metrics stay around the same level for both the random duty cycling and the genie-aided benchmarks. Meanwhile, the mean number of IoTDs with low-energy availability decreases with the device density for the KNN-based and RL-based proposals from around 67\% for 10 IoTDs to 8\%-13\% for 250 IoTDs. Similarly, the mean IoTD battery level per TTI increases from around 10 units to 72-86 units in high-density scenarios. For the optimized duty cycling proposal, the number of devices with low-energy availability fluctuates around 30\% - 45\% while the mean battery level is around 31-54 units. Moreover, for the proposal based on DT, the IoTDs with low energy availability also decrease in a 50\%, and the mean battery is always above 40 units per IoTD, increasing to almost 88 units per IoTD for high-density scenarios.    

Next, we analyze how many IoTDs are activated wrongly per event. This occurs when the activated IoTDs have low-energy availability and thus cannot transmit (failure) or when the BS wakes-up IoTDs far from the event epicenter. Fig.~\ref{wrong_fig}a shows this wrong activation of IoTDs while the corresponding energy level wasted due to the previous situations is shown in Fig.~\ref{wrong_fig}b. The genie-aided benchmark does not waste any energy since it always makes the best decision at first. Note that the energy wasted per IoTD decreases with the device density, but also more wrong activations occur. 
Notably, the optimized duty cycling leads to around 17 wrong activations per event for high-density scenarios while the RL-based proposal leads to 9.5 wrong IoTD activations pursuing a higher amount of information per event while the KNN-based proposal leads to less than 6 wrong activation. Using the DT-based proposal, the number of wrong IoTDs activations per event decreases by 50\%, outperforming the other proposed methods and lowering the energy wasted per event below 1.5 units per IoTD despite the device density, as shown in Fig.~\ref{wrong_fig}b).

\begin{figure}[t]
	\centering
	\includegraphics[width=0.9\columnwidth]{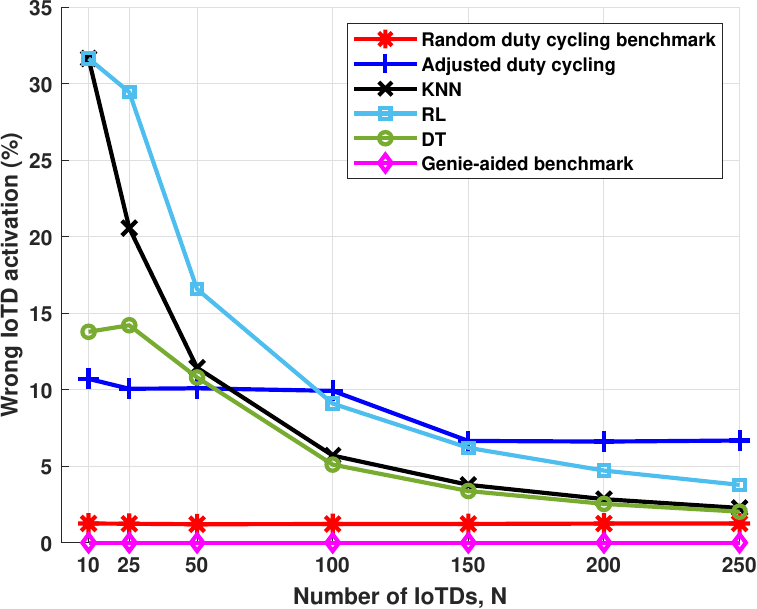}\vspace{-1.5ex}
        \includegraphics[width=0.9\columnwidth]{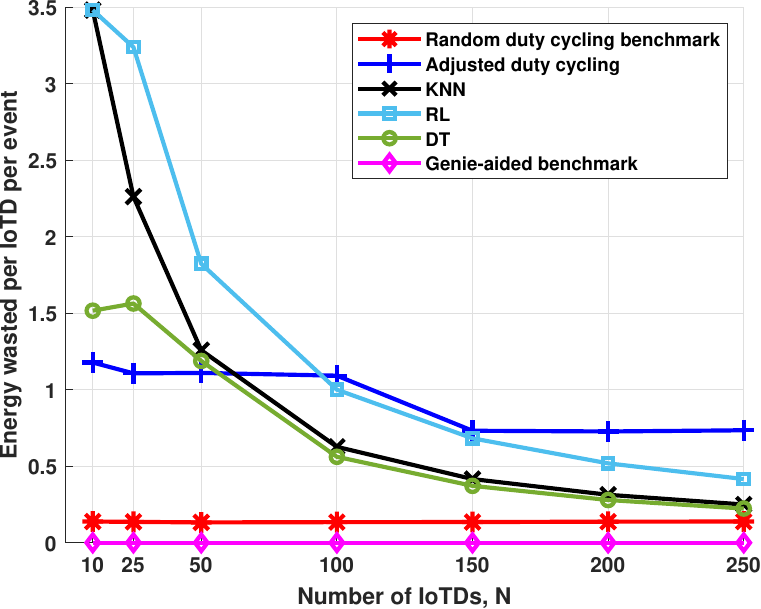}
  \vspace{-3mm}
    \caption{a) Wrong IoTD activation due to low-energy availability or event out of range (top), and b) the corresponding wasted energy as a function of the number of IoTDs (bottom).}
	\label{wrong_fig}
  \vspace{-6mm}
\end{figure}

Note that the proposed KNN-based and RL-based methods run on the BS, and the cost for the IoTDs is negligible. Moreover, the KNN-based proposal can be performed offline given the spatial information of the IoTDs. 

\vspace{-2mm}
\subsection{EH Analysis}
\vspace{-1mm}

The parameter $\lambda$ has a significant impact on the likelihood of the energy source being active within a TTI. A good understanding of $\lambda$ is crucial for efficient energy management in IoT systems, particularly those relying on EH. 
Fig.~\ref{EH_data} depicts the average energy units harvested per IoTD per TTI as the parameter $\lambda$ varies.  
The red region indicates the energy consumption range for the scenario with $N = 10$, while the green region corresponds to $N = 250$. Specifically, the upper/lower bounds of the red (green) regions correspond to the energy consumption when using the RL-based proposal/random duty cycling benchmark (adjusted duty cycling/DT-based proposal), as shown in Fig.~\ref{EC_fig}. 
It is important to note that when the IoTDs can harvest just one energy unit per energy arrival, the energy consumption is lower than the energy arrival per TTI when using the benchmark and proposals for $\lambda$ above 0.15-0.3 in high-density scenarios, whereas, for low-density scenarios, this increases to 0.4-0.9. In the case of 5 and 10 energy units per energy arrival, $\lambda$ could be as low as 0.05 and 0.02, respectively, for high-density scenarios, while for low-density scenarios, it increases to 0.09 and 0.05, respectively.  
\begin{figure}[t]
	\centering	\includegraphics[width=\columnwidth]{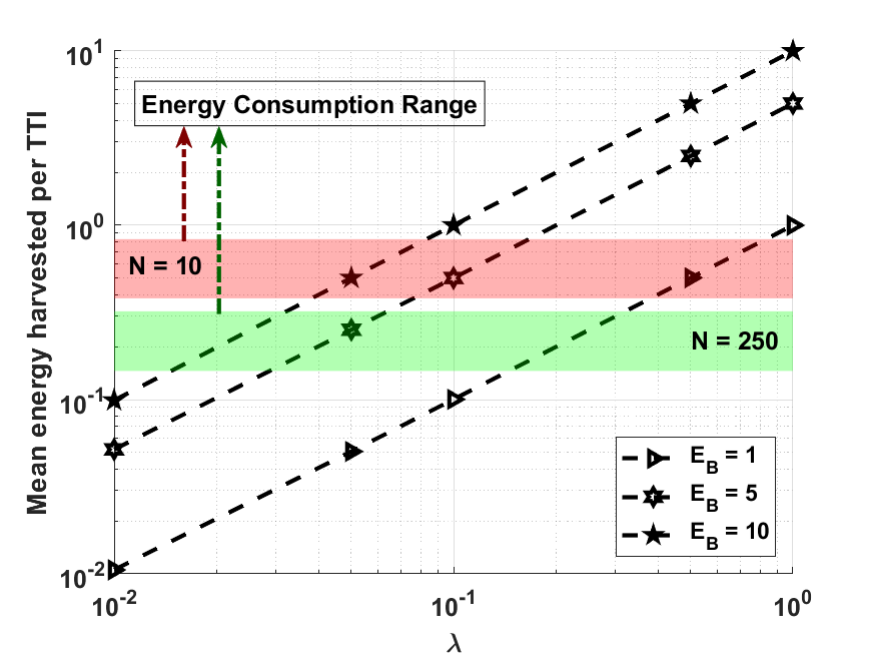}
  \vspace{-8mm}
    \caption{Mean energy units harvested per IoTD per TTI as a function of the EH rate ($\lambda$) for  $E_\text{B} \in\{1,5,10\}$ energy units. Light red and green areas show the mean energy consumption when using the benchmark and proposals for $N=10$ and $N=250$ respectively. Upper and lower bounds of the regions correspond respectively to the proposed RL and random benchmark methods for $N=10$ and adjusted duty cycling and DT methods for $N=250$.}
	\label{EH_data}
  \vspace{-3mm}
\end{figure}

Moreover, Fig.~\ref{EH_net} illustrates the net energy harvested per IoTD per TTI as $\lambda$ varies. The curves represent the methods with the most/least energy consumption adjusted duty cycling/random benchmark for $N=100$, as shown in Fig.~\ref{EC_fig}. We calculate the net energy harvested as the difference between the mean energy harvested and the energy consumption. Note that when IoTDs can harvest just one energy unit per energy arrival, the net energy harvested is positive for $\lambda > 0.38$ and $\lambda > 0.23$ for the upper/lower energy consumption bound, respectively. In the case of 5 and 10 energy units per energy arrival, $\lambda$ could be as low as 0.052/0.096 and 0.023/0.048, for the upper/lower bound, respectively.  

\section{Conclusions}\label{conclusion}

Energy-efficient strategies and optimized resource allocation are vital for the sustainability of IoT networks, especially those powered by EH. This work focused on an EH-driven IoT system for event monitoring, where IoTDs operate solely on harvested energy. IoTD behavior is modeled using a four-state Markov chain for sensing and sleep transitions, a modulated Poisson process for EH dynamics, and a discrete-time Markov chain for battery evolution. We proposed a spatially aware KNN-based duty cycling combined with WuR to reduce energy consumption while preserving detection accuracy. Additionally, RL and DT-based methods are introduced to maximize event-related information collection under energy constraints. Compared to a random duty cycling benchmark, our methods improve information received per event by up to 4.5 times and reduce energy use by up to 23\%, approaching the performance of an ideal genie-aided benchmark. 
These results highlight the effectiveness of combining spatial correlation-aware heuristics with ML for self-sustainable EH-IoT networks. Future work will address scalability and adaptive strategies for heterogeneous EH conditions. 

\begin{figure}[t]
	\centering	\includegraphics[width=0.95\columnwidth]{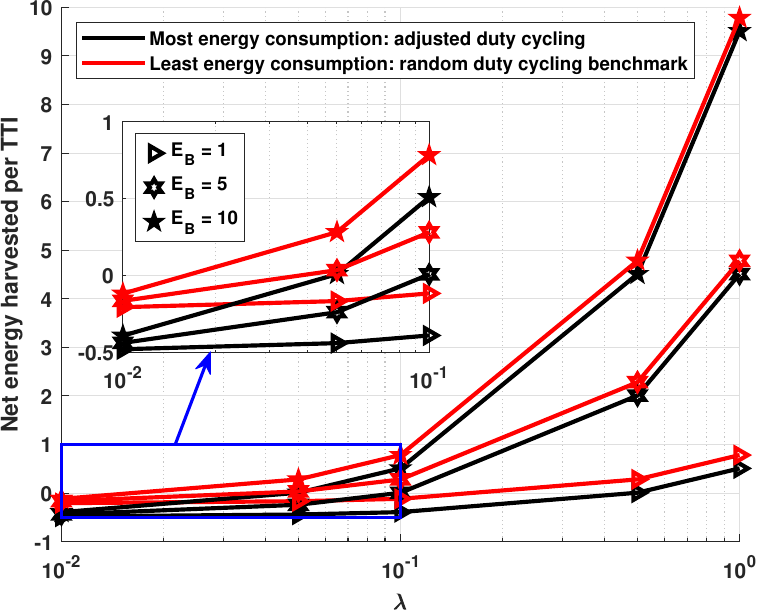}
  \vspace{-3mm}
    \caption{Mean net energy units harvested per IoTD per TTI as a function of the energy arrival rate ($\lambda$) for $E_\text{B} \in\{1,5,10\}$ energy units. The curves represent the upper/lower energy consumption bounds, which correspond to adjusted duty cycling/random benchmark for $N=100$.}
	\label{EH_net}
  \vspace{-3mm}
\end{figure}



\vspace{-2mm}

\bibliographystyle{IEEEtran}
\bibliography{short_bib}

\clearpage







\end{document}